\documentclass[aps,prl,reprint,superscriptaddress]{revtex4-1}
\pdfoutput=1
\usepackage{color}
\usepackage[normalem]{ulem}
\usepackage{graphicx}
\usepackage{amsmath}
\usepackage{amsfonts}
\usepackage{amssymb}
\usepackage{microtype}
\usepackage{verbatim}
\usepackage{bbold}
\usepackage{hyperref}
\allowdisplaybreaks
\usepackage{multirow}
\usepackage{bbm}
\usepackage{braket}
\usepackage{soul}
\usepackage{placeins} 
\hypersetup{
  colorlinks   = true, 
  urlcolor     = blue, 
  linkcolor    = blue, 
  citecolor   = blue 
}

\usepackage{textcomp}

\usepackage[smallerops]{newtxmath}

\begin{document}

\author{Tamiro Villazon}
\affiliation{Department of Physics, Boston University, 590 Commonwealth Ave., Boston, MA 02215, USA}

\author{Anushya Chandran}
\affiliation{Department of Physics, Boston University, 590 Commonwealth Ave., Boston, MA 02215, USA}

\author{Pieter W. Claeys}
\email{pc652@cam.ac.uk}
\affiliation{TCM Group, Cavendish Laboratory, University of Cambridge, Cambridge CB3 0HE, UK}

\title{Integrability and dark  states in an anisotropic central spin model}

\begin{abstract}

Central spin models describe a variety of quantum systems in which a spin-1/2 qubit interacts with a bath of surrounding spins, as realized in quantum dots and defect centers in diamond. 
We show that the fully anisotropic central spin Hamiltonian with (XX) Heisenberg interactions is integrable. 
Building on the class of integrable Richardson-Gaudin models, we derive an extensive set of conserved quantities and obtain the exact eigenstates using the Bethe ansatz.
These states divide into two exponentially large classes: \emph{bright} states, where the qubit is entangled with the bath, and \emph{dark} states, where it is not. 
We discuss how dark states limit qubit-assisted spin bath polarization and provide a robust long-lived quantum memory for qubit states.
\end{abstract}

\maketitle

\begin{figure}[!ht]
\includegraphics[width=0.90\columnwidth]{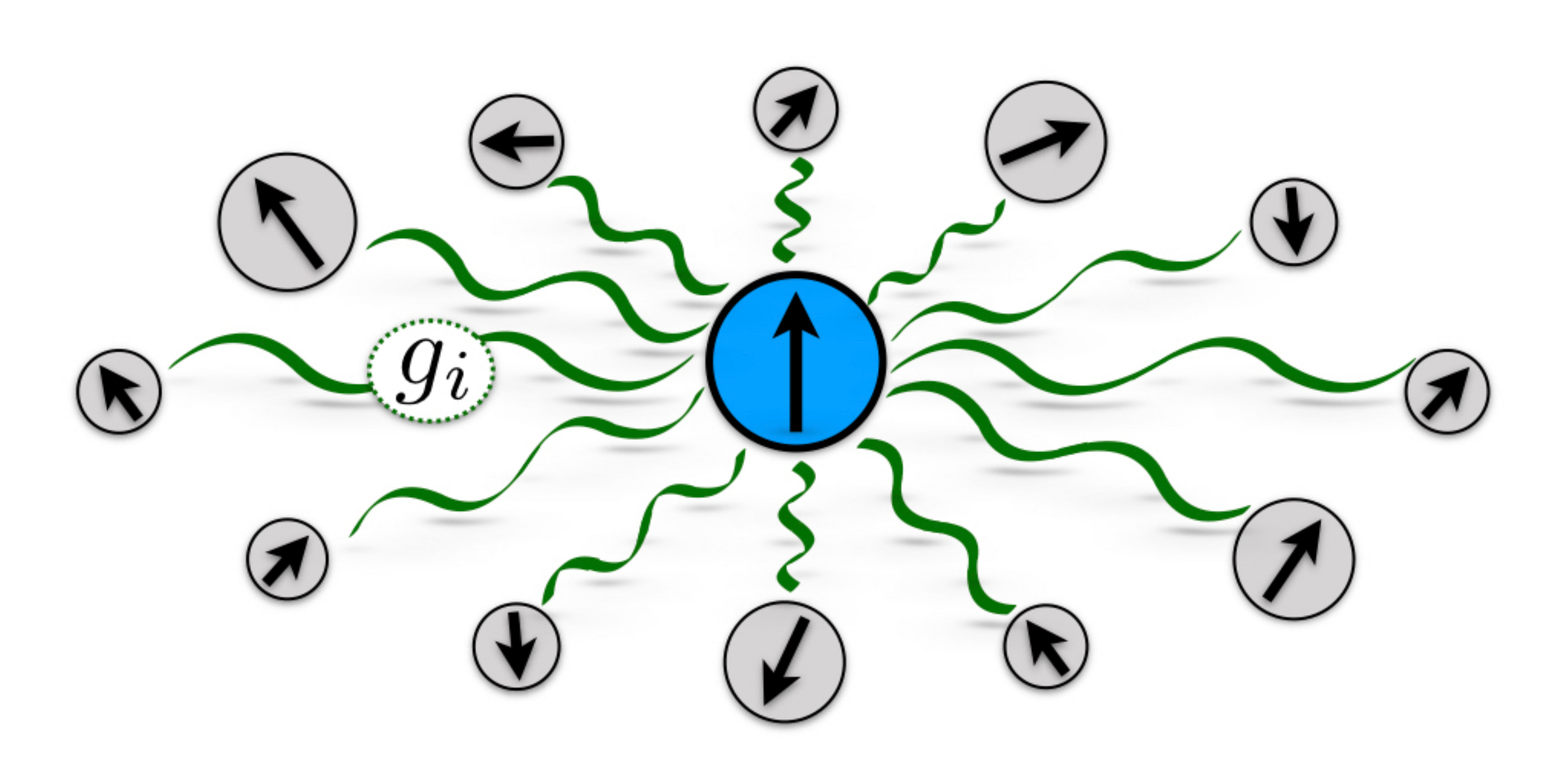}
\caption{Schematic representation of the central spin model. A central spin-$\frac{1}{2}$ particle interacts with an environment of $L$ spin-$s_i$ particles with interaction strengths $g_i$, $i=1 \dots L$. \label{fig:schematic}}
\vspace{0\baselineskip}
\end{figure}

\emph{Introduction.} -- With the advent of new quantum technologies, there is increasing interest in using small quantum systems to control and coherently manipulate mesoscopic environments \cite{kosloff_modeling_2019,koch_controlreview_2016,de_meso_2012,cai_simulator_2013,villazon_heat_2019,dong2019optimal}. In the simplest setting, a single spin-1/2 qubit controls a surrounding bath of spins, extending the available degrees of freedom and turning the detrimental effects of the bath into a useful resource. These systems are modelled by central spin (or spin star) Hamiltonians, as schematically illustrated in Fig.~\ref{fig:schematic}. Central spin models have broad applicability in quantum information~\cite{yung_networks_2011,tran_blind_2018}, quantum metrology and sensing~\cite{sushkov_magnetic_2014,he_exact_2019}, and describe the interactions between nitrogen-vacancy centers and nuclear spins in diamond~\cite{schwartz2018robust,london_polar_2013} and the hyperfine interaction in quantum dots~\cite{hanson_review_2007,schliemann2003electron,urbaszek2013nuclear}.

The central spin $\vec{S}_0$ typically interacts with the bath spins $\vec{S}_i$ through anisotropic Heisenberg interactions $\propto S_0^{x}S_i^{x} + S_0^{y} S_i^{y} + \alpha\, S_0^{z}S_i^{z}$. 
The fully isotropic XXX model ($\alpha = 1$) is common in systems with emergent spherical symmetry, e.g. quantum dots in semiconductors with $s$-type conduction bands~\cite{hanson_review_2007,schliemann2003electron}, while the fully anisotropic XX model ($\alpha = 0$) arises in resonant dipolar spin systems in rotating frames~\cite{hartmann1962nuclear,rovnyak2008tutorial,rao2019spin,lai2006knight,ding2014high,taylor2003long,fernandez-acebal_hyper_2017}.
Crucially, the fully isotropic XXX model is integrable, belonging to the class of XXX Richardson-Gaudin integrable models~\cite{gaudin_bethe_2014,dukelsky_colloquium_2004,rombouts_quantum_2010}. Integrability guarantees an extensive set of conserved quantities and allows all eigenstates to be exactly obtained using Bethe ansatz techniques, which has led to various studies of the equilibrium and dynamical properties of the XXX model~\cite{bortz_exact_2007,faribault_quantum_2009,bortz_dynamics_2010,faribault_integrability-based_2013,claeys_spin_2017,he_exact_2019}. However, generic central spin models with $\alpha \neq 1$ are not known to be integrable.

In this work, we show that the fully anisotropic XX model ($\alpha=0$) is integrable and exhibits a rich eigenstate structure. Its Hamiltonian describes a central spin-$\frac{1}{2}$ qubit in an external magnetic field $\omega_0$, interacting with a bath of $L$ spins in a uniform field $\omega$,
\begin{equation}\label{eq:CentralSpinHam}
H = \omega_0 S_0^z + \omega \sum_{i=1}^L S_i^z + \sum_{i=1}^L g_i \left(S_0^{+}S_i^- + S_0^-S_i^+\right).
\end{equation}
Here, the interaction strengths $g_i$ are taken to be inhomogeneous and the bath spins can have arbitrary spin $s_i$ (see Fig.~\ref{fig:schematic}). Since $H$ conserves total spin projection $S^z = S_0^z + \sum_{i} S_i^z$, we set $\omega=0$ without loss of generality. To establish the integrability of $H$, we present an extensive number of conserved charges and construct the exact eigenstates using the Bethe ansatz~\cite{gaudin_bethe_2014}. We note that Jivulescu \emph{et al.} previously used Bethe ansatz techniques to construct a subset of exact eigenstates~\cite{jivulescu_exact_2009,jivulescu_dynamical_2009}, but did not show that $H$ is integrable.

\begin{figure}[hb!]
\includegraphics[width=0.90\columnwidth]{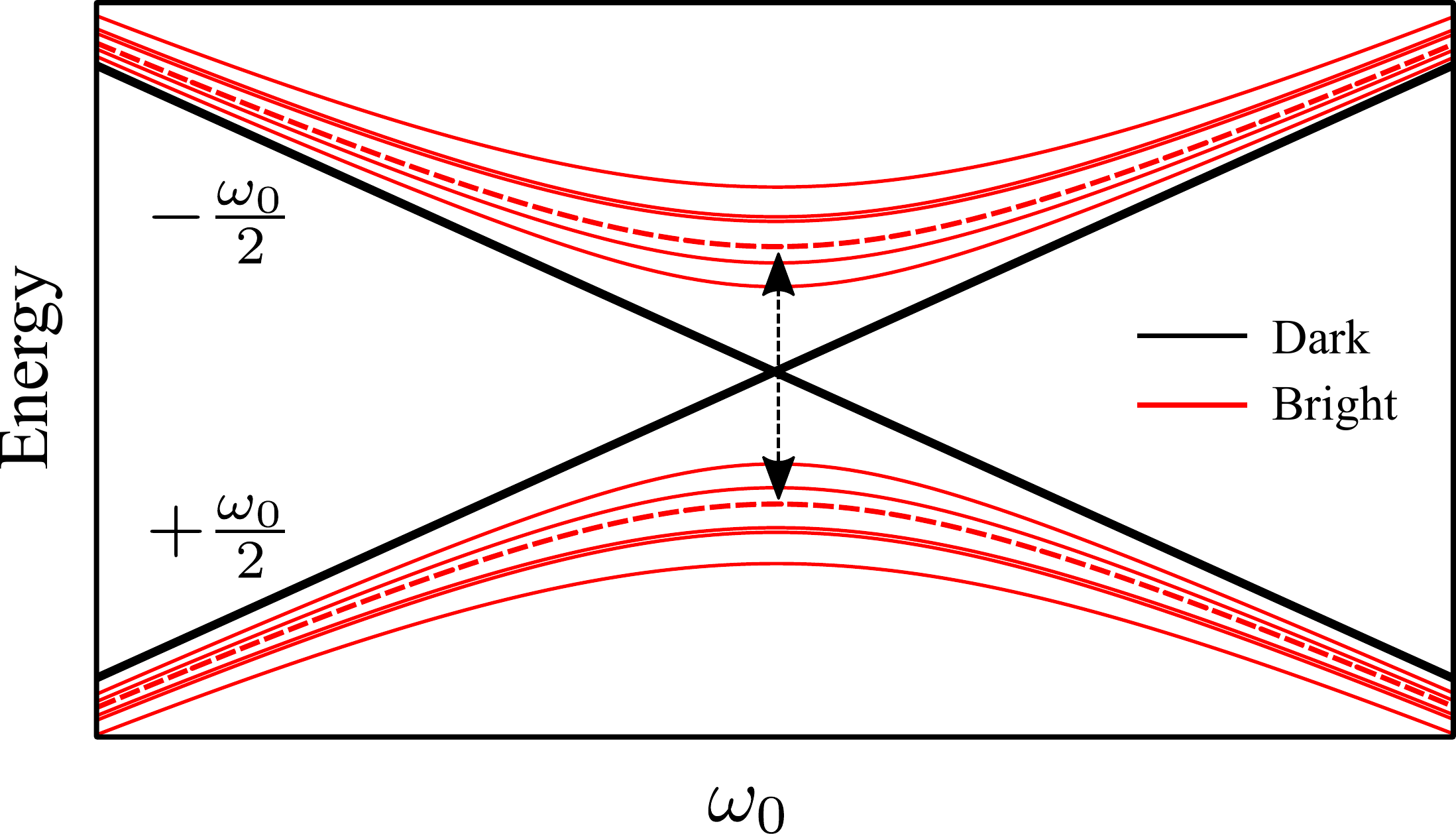}
\caption{Schematic representation of the energy spectrum as a function of the central field $\omega_0$, exhibiting highly-degenerate dark states (black lines) and bands of bright states (red lines). The dashed lines highlight an example pair of bright states exhibiting level repulsion near resonance $\omega_0 = \omega=0$.}
\label{fig:spectrum}
\vspace{-1\baselineskip}
\end{figure}

Remarkably, the eigenstates of the XX model separate into two exponentially large classes with distinct entanglement structure. \emph{Dark states} $\ket{\mathcal{D}}$, have a product state (unentangled) structure $\ket{\downarrow}_0 \otimes \ket{\mathcal{D}^-}$ or $\ket{\uparrow}_0 \otimes \ket{\mathcal{D}^+}$, where the central spin is fully polarized along the $z$-direction and the bath state satisfies
\begin{equation}\label{eq:DarkStates}
\left(\sum_{i=1}^L g_i S_i^{\pm} \right)\ket{\mathcal{D}^{\pm}} = 0.
\end{equation}
Dark states are independent of $\omega_0$ and form degenerate manifolds with energy $\pm \omega_0/2$ in every $S^z$ sector.
In contrast, \emph{bright states} $\ket{\mathcal{B}}$ exhibit qubit-bath entanglement and are given by linear combinations of definite central spin projection $c_{\downarrow} \ket{\downarrow}_0 \otimes \ket{\mathcal{B}^-} + c_{\uparrow} \ket{\uparrow}_0 \otimes \ket{\mathcal{B}^+}$. These states explicitly depend on $\omega_0$, and arise in pairs exhibiting level repulsion in the eigenspectrum of $H$ (schematically shown in Fig.~\ref{fig:spectrum}). 

Bright and dark states play an important role in qubit-assisted polarization of the spin bath ~\cite{taylor_controlling_2003,imamoglu_optical_2003,christ_nuclear_2009,belthangady2013dressed}.
Dark states have also been proposed as candidates for quantum memory, as their unentangled structure can be used to robustly store qubit states~\cite{taylor_controlling_2003,kurucz2009qubit}. We discuss the polarization and memory applications of the XX model in the concluding section.

\emph{Conserved charges.} -- The conserved charges of the Hamiltonian $H$ in  Eq.~\eqref{eq:CentralSpinHam} follow from the class of integrable XXZ Richardson-Gaudin models \cite{gaudin_bethe_2014,dukelsky_colloquium_2004,rombouts_quantum_2010} (see the explicit derivation below), and are given by
\onecolumngrid
\begin{align}\label{eq:defQi}
Q_i &= - 2 S_0^z S_i^z + \frac{1}{2}\left(S_i^+ S_i^- + S_i^- S_i^+\right)+ \sum_{\substack{j =1 \\ j \neq i}}^L\left[ \frac{g_i g_j}{g_i^2-g_j^2} \left(S_i^{+}S_j^- + S_i^-S_j^+\right)+ \frac{2\, g_j^2}{g_i^2-g_j^2} S_i^z S_j^z\right], \qquad i=1 \dots L. 
\end{align}
\twocolumngrid
These satisfy $[H,Q_i] = 0$ and $[Q_i,Q_j]=0, \forall i,j=1 \dots L$. Interpreting $H$ as $Q_0$, the number of linearly-independent conserved charges exactly equals the number of spins in the system. An additional conserved charge that helps establish integrability is given by
\begin{equation}
\tilde{Q} = - 2 S_0^z \sum_{i=1}^L g_i^2 S_i^z  + \frac{1}{2} \sum_{i,j=1}^L g_i g_j \left(S_i^{+}S_j^- + S_i^-S_j^+\right),
\end{equation}
which can be obtained as a linear combination of the conserved charges as $\tilde{Q} = \sum_{i=1}^L g_i^2 Q_i$. The charge $\tilde{Q}$ relates directly to the square of the Hamiltonian~
\begin{equation}\label{eq:QH2}
\tilde{Q} = H^2 - \omega_0^2/4,
\end{equation}
using the spin-$\frac{1}{2}$ properties of the central spin. Such quadratic relations between conserved charges are prevalent in Richardson-Gaudin systems~\cite{claeys_inner_2017,claeys_eigenvalue-based_2015,faribault_gaudin_2011,links_completeness_2016}. 

While the eigenstates of $\tilde{Q}$ at finite (non-zero) energy are not necessarily eigenstates of $H$, the zero-energy eigenstates of $\tilde{Q}$ are guaranteed to be eigenstates of $H$ with eigenvalue $\pm \omega_0/2$. 
It is precisely these zero modes that are the dark states. Eigenstates of $\tilde{Q}$ with non-zero eigenvalue $\Delta^2$ will be shown to be doubly-degenerate. In the spectrum of $H$, these states exhibit level repulsion, and split into a pair of bright states with energies $\pm \sqrt{\omega_0^2/4+\Delta^2}$ (see Fig.~\ref{fig:spectrum}).

\emph{Dark and bright states.} -- All eigenstates of $H$ can be expressed in terms of generalized spin raising operators acting on a vacuum state. To distinguish between dark and bright states, we assume $S^z<0$ in the main text and take the vacuum to be the state with all spins maximally down~\footnote{Any eigenstate can be constructed in two ways, either by expressing the state in terms of generalized spin raising operators acting on a vacuum state with all spins maximally down, or in terms of spin lowering operators acting on a dual vacuum state with all spins maximally up. For $S^z<0$, the former is more transparent and highlights the differences between bright states and dark states (with central spin necessarily down), and for $S^z>0$ the latter is more convenient since all dark states then have central spin up.}. 
Consider the Bethe states with $N<L/2$ spin excitations on top of the vacuum state $\ket{0} = \ket{\downarrow}_0 \otimes_{i=1}^L \ket{-s_i}$, 
\begin{equation}\label{eq:BetheStates}
\ket{\psi(v_1, v_2, \dots v_N)} = G^+(v_1)G^+(v_2) \dots G^+(v_N)\ket{0}, 
\end{equation}
with generalized spin raising operators that depend on (possibly complex) parameters $v_1, v_2, \dots v_N$ as
\begin{equation}
G^+(v) = \sum_{i=1}^L \frac{g_i}{1- g_i^2\ v}S_i^+.
\end{equation}
As no spin raising operators act on the central spin, the central spin points along $-z$ in the Bethe states. 

In the regime $S^z \leq 0$, the only allowed \emph{dark states} are those with central spin down, which are exactly of the form of Eq.~\eqref{eq:BetheStates}. Namely, dark states satisfy
\begin{align}\label{eq:BetheDarkState}
H\ket{\mathcal{D}}= -\frac{\omega_0}{2}\ket{\mathcal{D}}, \quad\quad \ket{\mathcal{D}}=\ket{\psi(v_1, v_2, \dots v_N)},
\end{align}
provided the rapidities $v_1 \dots v_N$ satisfy the Bethe equations 
\begin{equation}\label{eq:BetheDarkEq}
 \sum_{i=1}^L \frac{s_i g_i^2}{1- g_i^2\ v_a}-\sum_{b \neq a}^N \frac{1}{v_b-v_a} = 0,
\end{equation}
for $a=1 \dots N$. Importantly, these rapidities, and hence the structure of dark states, only depend on $\{g_i\}$ and not on $\omega_0$.

The \emph{bright states}, on the other hand, can be written as a linear combination of two Bethe states of the form \eqref{eq:BetheStates}, with central spin down and up respectively,
\begin{align}\label{eq:BetheBrightState}
&\ket{\phi(E; v_1, v_2, \dots v_{N-1})} =  \ket{\psi(0,v_1, v_2, \dots, v_{N-1})} \nonumber\\
&\qquad \qquad \qquad +\left(E+ \frac{\omega_0}{2}\right) S_0^+ \ket{\psi(v_1, v_2, \dots, v_{N-1})} \nonumber\\
&\qquad =(G^+(0)+\left(E+ \frac{\omega_0}{2}\right) S_0^+)\ket{\psi(v_1, v_2, \dots, v_{N-1})},
\end{align}
with rapidities now satisfying
\begin{align}
 &E^2 = \frac{\omega_0^2}{4}+ \left(\sum_{i=1}^L 2 s_i g_i^2 - \sum_{a=1}^{N-1} \frac{2}{v_a}\right),\label{eq:BetheBrightEq_Quad} \\
 &1+v_a \sum_{i=1}^L \frac{s_i g_i^2}{1-g_i^2\ v_a} - v_a \sum_{b \neq a}^{N-1} \frac{1}{v_b-v_a} =0, \label{eq:BetheBrightEq}
\end{align}
for $a=1, \dots , N-1$. The bright states satisfy
\begin{equation}
H\ket{\mathcal{B}}  = E \ket{\mathcal{B}}, \quad\quad \ket{\mathcal{B}}=\ket{\phi(E; v_1,  \dots, v_{N-1})}.
\end{equation}
While the rapidities $v_1, \dots v_{N-1}$ do not depend on $\omega_0$, the quadratic equation for the energy $E$ does. As such, each solution for $v_1 \dots v_{N-1}$ leads to two possible solutions for $E$, which exhibit level repulsion. These states are similar to nested Bethe ansatz states, with two different kinds of raising operators and two kinds of Bethe equations \cite{essler_one-dimensional_2005} .

\begin{figure}
\includegraphics[width=\columnwidth]{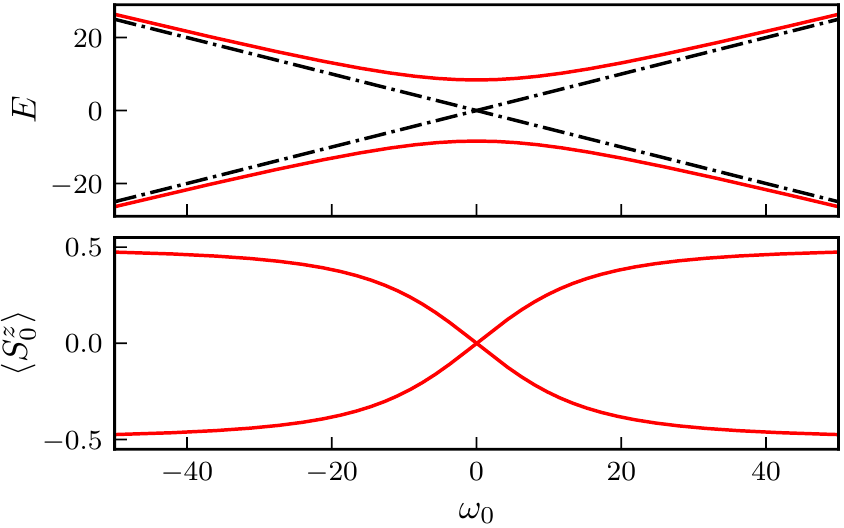}
\caption{Relation between energy $E$ and central spin expectation value $\braket{S_0^z}$ for the pair of bright states corresponding to the ground state and the highest excited state for a central spin system with $L=100$ and $N=L/4$, where we choose $(g_i)^{-2}=i$ and $s_i=1/2,\forall i= 1 \dots L$. Dashed lines denote $E=\pm \omega_0/2$. Full lines follow from $\sum_{i} 2s_i (g_i)^2 - \sum_{a} 2/v_a \approx 69.9787$, where the Bethe equations have been solved using the numerical methods developed in Ref.~\cite{claeys_eigenvalue-based_2015}.\label{fig:groundstate}}
\vspace{-\baselineskip}
\end{figure}

Interestingly, the expectation value of the central spin polarization of any eigenstate can be related to its energy $E$ by combining the Hellmann-Feynman theorem with Eq.~\eqref{eq:BetheBrightEq_Quad}:
\begin{equation}
\braket{S_0^z} = \left\langle \frac{\partial H}{\partial \omega_0}\right \rangle = \frac{\partial E}{\partial \omega_0} = \frac{1}{4}\frac{\omega_0}{E}.
\end{equation}
This relation is illustrated in Fig.~\ref{fig:groundstate} for a pair of bright states. For dark states, taking $E = \pm \omega_0/2$ returns the expected value of $\langle S_0^z \rangle = \pm 1/2$.

We end this section with a few comments. For $S^z<0$, the solutions to Eqs.~\eqref{eq:BetheDarkEq} and ~\eqref{eq:BetheBrightEq} respectively identify dark states in which the central spin points along $-z$, and bright states. For $S^z=0$, the Bethe equations~\eqref{eq:BetheDarkEq} do not admit any solutions, and consequently, all the states in the spectrum are bright.
More generally, the Bethe equations do not admit any solutions if $S^z \geq 0$.
Nevertheless, the spectrum for $S^z > 0$ has bright and dark states, with the central spin along $+z$ in the dark states.
The dark states now arise as additional solutions to the Bethe equations in Eq.~\eqref{eq:BetheBrightEq} in which some of the rapidities are zero. They are obtained by taking the limit $E\to \infty$ in Eq.~\eqref{eq:BetheBrightState} (as a zero rapidity implies $E \to \infty$ in Eq. \eqref{eq:BetheBrightEq_Quad}). In this regime, dark state energies are not given by $E$, but by $+\omega_0/2$.

\emph{Explicit derivation}. -- The structure of the conserved charges can be easily understood starting from $\tilde{Q}$ rather than $H$. To this end, we rewrite $\tilde{Q}$ as
\begin{align}\label{eq:defQ}
 \tilde{Q} =& \left(\frac{1}{2}+S_0^z\right) \left( G^- G^+ \right) \left(\frac{1}{2}+S_0^z\right) \nonumber \\
 &\qquad + \left(\frac{1}{2}-S_0^z\right) \left(G^+ G^-\right) \left(\frac{1}{2}-S_0^z\right).
\end{align}
Above, $G^{\pm} \equiv G^{\pm}(0) = \sum_{i=1}^L g_i S_i^{\pm}$, and $(1/2 \pm S_0^z)$ projects the central spin along the $z$-direction. Hamiltonians of the form $G^\pm G^\mp$ arise in the study of topological superconductivity and superfluidity \cite{ibanez_exactly_2009,lerma_h._integrable_2011,ortiz_many-body_2014}, neutron pairing \cite{dean_pairing_2003}, and Sachdev-Ye-Kitaev-like models \cite{iyoda_effective_2018}. These are Richardson-Gaudin integrable, and all results for their eigenstates and conserved charges can be found in, e.g., Refs.~\cite{dukelsky_class_2001,dukelsky_colloquium_2004,ortiz_exactly-solvable_2005,rombouts_quantum_2010,van_raemdonck_exact_2014,links_exact_2015,iyoda_effective_2018}.

The conserved charges of  $G^- G^+$ are known to be
\begin{align}
Q_i^{(-+)} &= S_i^-S_i^+ + 2 \sum_{j \neq i}^L \frac{g_j^2}{g_i^2-g_j^2} S_i^z S_j^z \nonumber \\
&\qquad +\sum_{j \neq i}^L \frac{g_i g_j}{g_i^2-g_j^2}\left(S_i^+ S_j^-+S_i^- S_j^+\right),
\end{align}
which mutually commute and commute with $G^-G^+$. The conserved charges $Q_i^{(+-)}$of $G^+G^-$ immediately follow from those of $G^-G^+$ through spin inversion symmetry, since mapping $S^{\pm} \to S^{\mp}$ and $S^z \to -S^z$ leaves the $su(2)$ algebra invariant and maps $G^- G^+ \to G^+ G^-$. The conserved quantities of $\tilde{Q}$ then follow by combining these with the appropriate projection operators from Eq.~\eqref{eq:defQ} as
\begin{align}
Q_i &= \left(\frac{1}{2}+S_0^z\right) Q_i^{(-+)}  \left(\frac{1}{2}+S_0^z\right) \nonumber \\
 &\qquad + \left(\frac{1}{2}-S_0^z\right) Q_i^{(+-)} \left(\frac{1}{2}-S_0^z\right). \label{Eq:Qi-Qipm}
 \end{align}
Using the spin-$1/2$ properties of the central spin, Eq.~\eqref{Eq:Qi-Qipm} simplifies to Eq.~\eqref{eq:defQi}. It can be easily checked that, not only do these conserved charges satisfy $[Q_i,Q_j]=0, \forall i,j$, but they also satisfy $[H,Q_i]=0, \forall i$.

The eigenstates of $H$ can be similarly derived by making the connection with the eigenstates of $\tilde{Q}$. The eigenstates of $G^+ G^-$ can be written as two kinds of Bethe states of the form \eqref{eq:BetheStates}, depending on the number of zero rapidities (see e.g. Ref.~\cite{iyoda_effective_2018}). The first kind are zero-energy eigenstates where all rapidities are nonzero and satisfy Eq.~\eqref{eq:BetheDarkEq},
\begin{equation} \label{eq:GG1}
G^+ G^- \ket{\psi(v_1, \dots, v_N)} = 0.
\end{equation}
The action of $\tilde{Q}$ reduces to the action of $G^+ G^-$ on these states because $S_0^z=-1/2$ in the vacuum state in Eq.~\eqref{eq:BetheStates}. As such, the zero-energy eigenstates of $G^+ G^-$ with central spin down are zero-energy eigenstates of $\tilde{Q}$, and are thus the dark states $|\mathcal{D}\rangle$. The condition $G^{-}|\mathcal{D}\rangle = 0$ in Eq.~\eqref{eq:DarkStates} follows directly from Eq.~\eqref{eq:GG1} since $G^+ G^-$ is positive definite. The zero modes of $\tilde{Q}$ with central spin up and their properties are similarly obtained from the zero modes of $G^- G^+$.

The second kind of eigenstates of $G^+ G^-$ have one zero rapidity and non-zero positive eigenvalues:
\begin{align}
&G^+ G^- \ket{\psi(0, v_1, \dots, v_{N-1})}  \nonumber\\
&\qquad= \left(\sum_{i=1}^L 2 s_i g_i^2 - \sum_{a=1}^{N-1}\frac{2}{v_a}\right) \ket{\psi(0, v_1, \dots, v_{N-1})}, \label{Eq:GpG-eig}
\end{align}
where the non-zero rapidities satisfy Eq.~\eqref{eq:BetheBrightEq}. 
Furthermore, it is easily shown that the state $\ket{\psi(v_1, \dots, v_{N-1})}$ is an eigenstate of $G^- G^+$ with the same eigenvalue as in Eq.~\eqref{Eq:GpG-eig}. Both states can be made into eigenstates of $\tilde{Q}$ by applying the proper central spin projectors. It follows that $\tilde{Q}$ has doubly-degenerate eigenstates 
\begin{align}\label{eq:degBrightStates}
G^+(0)\ket{\psi(v_1, \dots, v_{N-1})}, \ \ \ S_0^+ \ket{\psi(v_1, \dots, v_{N-1})},
\end{align}
with the same number of spin excitations $N$. Acting with $H$ on either of these states creates of linear superposition of them, and the eigenvalue equation for $H$ in this two-dimensional space leads to the quadratic equation \eqref{eq:BetheBrightEq_Quad} for the states \eqref{eq:BetheBrightState}. 

As can be expected, the degenerate pairs of states \eqref{eq:degBrightStates} are also eigenstates for each separate conserved charge $Q_i$. This follows from the observation that 
\begin{equation}
Q_i^{(+-)}G^+(0) = G^+(0) Q_i^{(-+)},
\end{equation}
such that if $\ket{\psi(v_1, \dots, v_{N-1})}$ is an eigenstate of $Q_i^{(-+)}$ then $G^+(0)\ket{\psi(v_1, \dots, v_{N-1})}$ is an eigenstate of $Q_i^{(+-)}$ with the same eigenvalue. The eigenvalues of $Q_i$ then follow from the known eigenvalues of the conserved charges of either $G^+ G^-$ or $G^- G^+$ (see e.g. Ref.~\cite{iyoda_effective_2018}). The eigenvalue $q_i$ of $Q_i$ for the dark states is given by
\begin{align}
q_i = 2 \,s_i &\left(\sum_{a=1}^N \frac{1}{1-g_i^2 \ v_a}  + \sum_{\substack{j=1 \\ j \neq i}}^L \frac{s_j  g_j^2}{g_i^2-g_j^2}  \right),
\end{align}
and for the bright states by
\begin{align}
q_i = 2\, s_i &\left(1+ \sum_{a=1}^{N-1} \frac{1}{1- g_i^2 \ v_a}  +  \sum_{\substack{j=1 \\ j \neq i}}^L \frac{s_j  g_j^2}{g_i^2-g_j^2}  \right).
\end{align}

\emph{Counting of states.} -- The number of dark states can be obtained in various ways, either using dimensionality arguments or by counting the number of possible solutions to the Bethe equations. We focus on a system with a bath of $L$ spin-$1/2$ particles and $S^z < 0$; the generalization to $S^z > 0$ and any choice of bath spins is straightforward. Dark states with $N<L/2$ spin excitations have central spin down and live within a ${L \choose N}$-dimensional subspace of the Hilbert space. By Eq.~\eqref{eq:DarkStates}, the dark states are projected by $G^{-}$ onto the zero-vector of the ${L \choose N-1}$-dimensional subspace of $N-1$ spin excitations. The corresponding dark state manifold thus consists of precisely those ${L \choose N} - {L \choose N-1}$ orthogonal states with no parallel projection. This is exactly the number of solutions to the Bethe equations \eqref{eq:BetheDarkEq} for dark states; the number of solutions to the Bethe equations are well-studied in the literature~\cite{links_exact_2015}. A similar expression for the number of dark states was presented in Ref.~\cite{taylor_controlling_2003}.

For bright states, the number of distinct solutions to Eq.~\eqref{eq:BetheBrightEq} is given by ${L \choose N-1}$, and each solution for the set $v_1, v_2, \dots, v_{N-1}$ leads to two possible solutions for $E$ in Eq. \eqref{eq:BetheBrightEq_Quad} and a pair of bright states. Combined, the total number of eigenstates is given by 
\begin{equation}
 \left({L \choose N} - {L \choose N-1}\right) + 2 {L \choose N-1} = {L+1 \choose N},
\end{equation}
returning the expected number of eigenstates in each $S^z$ sector and showing the expected completeness of the Bethe ansatz for spin-$\frac{1}{2}$ Richardson-Gaudin systems \cite{links_completeness_2016}. It also follows from the scaling of the binomial coefficients that the number of both dark and bright states grows exponentially with bath size $L$.

\emph{Discussion.} -- We established a new family of integrable Richardson-Gaudin central spin models with anisotropic XX interactions by deriving the full set of conserved charges and Bethe eigenstates. The eigenstates can be divided in two classes, dark or bright, depending on their qubit-bath entanglement properties. 

Dark states exhibit no qubit-bath entanglement and can be used to store qubit states for quantum memory. 
Ref.~\cite{taylor_controlling_2003} proposed a scheme to store (and retrieve) an arbitrary qubit state in the state of the bath using adiabatic passage in the XXX model with weak inhomogeneous couplings $g_j$. This scheme immediately generalizes to the XX model, even in the presence of strong inhomogeneities. 
Starting from a product state in which the central spin is in the desired qubit state and the environment in a dark state, the qubit state can be encoded in a superposition of dark and bright states (taking e.g. large positive $\omega_0$ values in Eq.~\eqref{eq:BetheBrightState}).
Since the dark component of the wavefunction is independent of $\omega_0$, an adiabatic passage to large negative values of $\omega_0$ transfers the qubit state to the bath as
\begin{align} \nonumber
&\left( u \ket{\downarrow}_0 + v \ket{\uparrow}_0 \right) \otimes \ket{\mathcal{D}^-} \\ \nonumber
& = u \ket{\downarrow}_0 \otimes \ket{\mathcal{D}^-}  + v \ket{\uparrow}_0  \otimes \sum_{\mathcal{B}^+}\ket{\mathcal{B}^{+}}\braket{\mathcal{B}^+|\mathcal{D}^-} \\ \nonumber
&  \xrightarrow{\omega_0\to-\infty}  \ket{\downarrow}_0 \otimes \left[u \, \ket{\mathcal{D}^-}  + v\,  G^+  \sum_{\mathcal{B}^{+}} e^{i\phi_{\mathcal{B}}}\ket{\mathcal{B}^{+}}\braket{\mathcal{B}^{+}|\mathcal{D}^-}\right] 
\end{align}
where $\phi_{\mathcal{B}}$ are the relative phases accrued during the passage. 
The final bath state serves as a robust memory for the qubit state; the qubit state can be retrieved by symmetrically reversing the process and accounting for the phases $\phi_\mathcal{B}$. 
This invites quantum memory applications in defect center or dot systems which are well described by the XX model.

Dark states are also known to limit hyperpolarization protocols which use a central qubit to transfer polarization to/from a bath~\cite{urbaszek2013nuclear}. In such protocols, the qubit is repeatedly polarized and manipulated (e.g. by tuning the field $\omega_0$) to induce qubit-bath exchange interactions. Since qubit-bath polarization exchanges are only possible in bright states, the bath polarization saturates to a value determined by the populated dark states. Several experiments have found saturation at high (above $60\%$) spin bath polarizations~\cite{bracker2005optical,urbaszek2007efficient,chekhovich2010pumping}, and strategies to overcome the limitations imposed by dark states have been proposed \cite{christ_nuclear_2009,imamoglu_optical_2003}. The explicit structure of dark states presented in our work may allow for the development of new hyperpolarization protocols with high saturation values of the bath polarization.

Not only the existence of bright and dark states, but also the integrability of the XX model has both physical and theoretical implications. 
Physically, integrable models are known to exhibit non-ergodic behaviour at long times and long-lived correlations due to the presence of conservation laws~\cite{jaynes1957information,rigol2007relaxation,vidmar2016generalized}. 
For the XX model, our expression for the conserved quantities $Q_i$ in Eq.~\eqref{eq:defQi} suggests that observables such as $S_0^z S_i^z$ may not thermalize at long times.
Theoretically, the Bethe ansatz allows for exact theoretical and numerical studies in system sizes beyond the reach of exact diagonalization~\cite{dukelsky_colloquium_2004,faribault_exact_2008,foster_quantum_2013,faribault_integrability-based_2013,el_araby_order_2014,claeys_spin_2017,james_non-perturbative_2018,robinson_computing_2019}, providing new avenues for the study of non-equilibrium dynamics in central spin models. 

\emph{Acknowledgements.} -- We thank Anatoli Polkovnikov and Stijn De Baerdemacker for useful discussions. P.W.C. gratefully acknowledges support from a Francqui Foundation Fellowship from the Belgian American Educational Foundation (BAEF), Boston University's Condensed Matter Theory Visitors program, and EPSRC Grant No. EP/P034616/1.  A.C. acknowledges support from the Sloan Foundation through Sloan Research Fellowships. This work was supported by NSF DMR-1813499 (T.V.), and NSF DMR-1752759 (T.V. and A.C.).

\bibliography{IntegrabilityCentralSpin}

\begin{thebibliography}{63}%
\makeatletter
\providecommand \@ifxundefined [1]{%
 \@ifx{#1\undefined}
}%
\providecommand \@ifnum [1]{%
 \ifnum #1\expandafter \@firstoftwo
 \else \expandafter \@secondoftwo
 \fi
}%
\providecommand \@ifx [1]{%
 \ifx #1\expandafter \@firstoftwo
 \else \expandafter \@secondoftwo
 \fi
}%
\providecommand \natexlab [1]{#1}%
\providecommand \enquote  [1]{``#1''}%
\providecommand \bibnamefont  [1]{#1}%
\providecommand \bibfnamefont [1]{#1}%
\providecommand \citenamefont [1]{#1}%
\providecommand \href@noop [0]{\@secondoftwo}%
\providecommand \href [0]{\begingroup \@sanitize@url \@href}%
\providecommand \@href[1]{\@@startlink{#1}\@@href}%
\providecommand \@@href[1]{\endgroup#1\@@endlink}%
\providecommand \@sanitize@url [0]{\catcode `\\12\catcode `\$12\catcode
  `\&12\catcode `\#12\catcode `\^12\catcode `\_12\catcode `\%12\relax}%
\providecommand \@@startlink[1]{}%
\providecommand \@@endlink[0]{}%
\providecommand \url  [0]{\begingroup\@sanitize@url \@url }%
\providecommand \@url [1]{\endgroup\@href {#1}{\urlprefix }}%
\providecommand \urlprefix  [0]{URL }%
\providecommand \Eprint [0]{\href }%
\providecommand \doibase [0]{https://doi.org/}%
\providecommand \selectlanguage [0]{\@gobble}%
\providecommand \bibinfo  [0]{\@secondoftwo}%
\providecommand \bibfield  [0]{\@secondoftwo}%
\providecommand \translation [1]{[#1]}%
\providecommand \BibitemOpen [0]{}%
\providecommand \bibitemStop [0]{}%
\providecommand \bibitemNoStop [0]{.\EOS\space}%
\providecommand \EOS [0]{\spacefactor3000\relax}%
\providecommand \BibitemShut  [1]{\csname bibitem#1\endcsname}%
\let\auto@bib@innerbib\@empty
\bibitem [{\citenamefont {Kosloff}(2019)}]{kosloff_modeling_2019}%
  \BibitemOpen
  \bibfield  {author} {\bibinfo {author} {\bibfnamefont {R.}~\bibnamefont
  {Kosloff}},\ }\bibfield  {title} {\bibinfo {title} {Quantum thermodynamics
  and open-systems modeling},\ }\href
  {https://doi.org/10.1063/1.5096173@jcp.2019.OSQD2019.issue-1} {\bibfield
  {journal} {\bibinfo  {journal} {J. Chem. Phys.}\ }\textbf {\bibinfo {volume}
  {150}},\ \bibinfo {pages} {204105} (\bibinfo {year} {2019})}\BibitemShut
  {NoStop}%
\bibitem [{\citenamefont {Koch}(2016)}]{koch_controlreview_2016}%
  \BibitemOpen
  \bibfield  {author} {\bibinfo {author} {\bibfnamefont {C.~P.}\ \bibnamefont
  {Koch}},\ }\bibfield  {title} {\bibinfo {title} {Controlling open quantum
  systems: Tools, achievements, and limitations},\ }\href
  {https://doi.org/10.1088/0953-8984/28/21/213001} {\bibfield  {journal}
  {\bibinfo  {journal} {J. Phys. Condens. Matter}\ }\textbf {\bibinfo {volume}
  {28}},\ \bibinfo {pages} {213001} (\bibinfo {year} {2016})}\BibitemShut
  {NoStop}%
\bibitem [{\citenamefont {De~Lange}\ \emph {et~al.}(2012)\citenamefont
  {De~Lange}, \citenamefont {Van Der~Sar}, \citenamefont {Blok}, \citenamefont
  {Wang}, \citenamefont {Dobrovitski},\ and\ \citenamefont
  {Hanson}}]{de_meso_2012}%
  \BibitemOpen
  \bibfield  {author} {\bibinfo {author} {\bibfnamefont {G.}~\bibnamefont
  {De~Lange}}, \bibinfo {author} {\bibfnamefont {T.}~\bibnamefont {Van
  Der~Sar}}, \bibinfo {author} {\bibfnamefont {M.}~\bibnamefont {Blok}},
  \bibinfo {author} {\bibfnamefont {Z.-H.}\ \bibnamefont {Wang}}, \bibinfo
  {author} {\bibfnamefont {V.}~\bibnamefont {Dobrovitski}},\ and\ \bibinfo
  {author} {\bibfnamefont {R.}~\bibnamefont {Hanson}},\ }\bibfield  {title}
  {\bibinfo {title} {Controlling the quantum dynamics of a mesoscopic spin bath
  in diamond},\ }\href {https://doi.org/10.1038/srep00382} {\bibfield
  {journal} {\bibinfo  {journal} {Sci. Rep.}\ }\textbf {\bibinfo {volume}
  {2}},\ \bibinfo {pages} {382} (\bibinfo {year} {2012})}\BibitemShut {NoStop}%
\bibitem [{\citenamefont {Cai}\ \emph {et~al.}(2013)\citenamefont {Cai},
  \citenamefont {Retzker}, \citenamefont {Jelezko},\ and\ \citenamefont
  {Plenio}}]{cai_simulator_2013}%
  \BibitemOpen
  \bibfield  {author} {\bibinfo {author} {\bibfnamefont {J.}~\bibnamefont
  {Cai}}, \bibinfo {author} {\bibfnamefont {A.}~\bibnamefont {Retzker}},
  \bibinfo {author} {\bibfnamefont {F.}~\bibnamefont {Jelezko}},\ and\ \bibinfo
  {author} {\bibfnamefont {M.~B.}\ \bibnamefont {Plenio}},\ }\bibfield  {title}
  {\bibinfo {title} {A large-scale quantum simulator on a diamond surface at
  room temperature},\ }\href {https://doi.org/10.1038/nphys2519} {\bibfield
  {journal} {\bibinfo  {journal} {Nat. Phys.}\ }\textbf {\bibinfo {volume}
  {9}},\ \bibinfo {pages} {168} (\bibinfo {year} {2013})}\BibitemShut {NoStop}%
\bibitem [{\citenamefont {Villazon}\ \emph {et~al.}(2019)\citenamefont
  {Villazon}, \citenamefont {Polkovnikov},\ and\ \citenamefont
  {Chandran}}]{villazon_heat_2019}%
  \BibitemOpen
  \bibfield  {author} {\bibinfo {author} {\bibfnamefont {T.}~\bibnamefont
  {Villazon}}, \bibinfo {author} {\bibfnamefont {A.}~\bibnamefont
  {Polkovnikov}},\ and\ \bibinfo {author} {\bibfnamefont {A.}~\bibnamefont
  {Chandran}},\ }\bibfield  {title} {\bibinfo {title} {Swift heat transfer by
  fast-forward driving in open quantum systems},\ }\href
  {https://doi.org/10.1103/PhysRevA.100.012126} {\bibfield  {journal} {\bibinfo
   {journal} {Phys. Rev. A}\ }\textbf {\bibinfo {volume} {100}},\ \bibinfo
  {pages} {012126} (\bibinfo {year} {2019})}\BibitemShut {NoStop}%
\bibitem [{\citenamefont {Dong}\ \emph {et~al.}(2019)\citenamefont {Dong},
  \citenamefont {Liang}, \citenamefont {Duan}, \citenamefont {Wang},
  \citenamefont {Li}, \citenamefont {Rong},\ and\ \citenamefont
  {Du}}]{dong2019optimal}%
  \BibitemOpen
  \bibfield  {author} {\bibinfo {author} {\bibfnamefont {L.}~\bibnamefont
  {Dong}}, \bibinfo {author} {\bibfnamefont {H.}~\bibnamefont {Liang}},
  \bibinfo {author} {\bibfnamefont {C.-K.}\ \bibnamefont {Duan}}, \bibinfo
  {author} {\bibfnamefont {Y.}~\bibnamefont {Wang}}, \bibinfo {author}
  {\bibfnamefont {Z.}~\bibnamefont {Li}}, \bibinfo {author} {\bibfnamefont
  {X.}~\bibnamefont {Rong}},\ and\ \bibinfo {author} {\bibfnamefont
  {J.}~\bibnamefont {Du}},\ }\bibfield  {title} {\bibinfo {title} {Optimal
  control of a spin bath},\ }\href
  {https://doi.org/https://doi.org/10.1103/PhysRevA.99.013426} {\bibfield
  {journal} {\bibinfo  {journal} {Phys. Rev. A}\ }\textbf {\bibinfo {volume}
  {99}},\ \bibinfo {pages} {013426} (\bibinfo {year} {2019})}\BibitemShut
  {NoStop}%
\bibitem [{\citenamefont {Yung}(2011)}]{yung_networks_2011}%
  \BibitemOpen
  \bibfield  {author} {\bibinfo {author} {\bibfnamefont {M.-H.}\ \bibnamefont
  {Yung}},\ }\bibfield  {title} {\bibinfo {title} {Spin star as a switch for
  quantum networks},\ }\href {https://doi.org/10.1088/0953-4075/44/13/135504}
  {\bibfield  {journal} {\bibinfo  {journal} {J. Phys. B: At., Mol. Opt.
  Phys.}\ }\textbf {\bibinfo {volume} {44}},\ \bibinfo {pages} {135504}
  (\bibinfo {year} {2011})}\BibitemShut {NoStop}%
\bibitem [{\citenamefont {Tran}\ and\ \citenamefont
  {Taylor}(2018)}]{tran_blind_2018}%
  \BibitemOpen
  \bibfield  {author} {\bibinfo {author} {\bibfnamefont {M.~C.}\ \bibnamefont
  {Tran}}\ and\ \bibinfo {author} {\bibfnamefont {J.~M.}\ \bibnamefont
  {Taylor}},\ }\bibfield  {title} {\bibinfo {title} {Blind quantum computation
  using the central spin {Hamiltonian}},\ }\href
  {http://arxiv.org/abs/1801.04006} {\bibfield  {journal} {\bibinfo  {journal}
  {arXiv:1801.04006 [quant-ph]}\ } (\bibinfo {year} {2018})}\BibitemShut
  {NoStop}%
\bibitem [{\citenamefont {Sushkov}\ \emph {et~al.}(2014)\citenamefont
  {Sushkov}, \citenamefont {Lovchinsky}, \citenamefont {Chisholm},
  \citenamefont {Walsworth}, \citenamefont {Park},\ and\ \citenamefont
  {Lukin}}]{sushkov_magnetic_2014}%
  \BibitemOpen
  \bibfield  {author} {\bibinfo {author} {\bibfnamefont {A.}~\bibnamefont
  {Sushkov}}, \bibinfo {author} {\bibfnamefont {I.}~\bibnamefont {Lovchinsky}},
  \bibinfo {author} {\bibfnamefont {N.}~\bibnamefont {Chisholm}}, \bibinfo
  {author} {\bibfnamefont {R.}~\bibnamefont {Walsworth}}, \bibinfo {author}
  {\bibfnamefont {H.}~\bibnamefont {Park}},\ and\ \bibinfo {author}
  {\bibfnamefont {M.}~\bibnamefont {Lukin}},\ }\bibfield  {title} {\bibinfo
  {title} {Magnetic {resonance} {detection} of {individual} {proton} {spins}
  {using} {quantum} {reporters}},\ }\href
  {https://doi.org/10.1103/PhysRevLett.113.197601} {\bibfield  {journal}
  {\bibinfo  {journal} {Phys. Rev. Lett.}\ }\textbf {\bibinfo {volume} {113}},\
  \bibinfo {pages} {197601} (\bibinfo {year} {2014})}\BibitemShut {NoStop}%
\bibitem [{\citenamefont {He}\ \emph {et~al.}(2019)\citenamefont {He},
  \citenamefont {Chesi}, \citenamefont {Lin},\ and\ \citenamefont
  {Guan}}]{he_exact_2019}%
  \BibitemOpen
  \bibfield  {author} {\bibinfo {author} {\bibfnamefont {W.-B.}\ \bibnamefont
  {He}}, \bibinfo {author} {\bibfnamefont {S.}~\bibnamefont {Chesi}}, \bibinfo
  {author} {\bibfnamefont {H.-Q.}\ \bibnamefont {Lin}},\ and\ \bibinfo {author}
  {\bibfnamefont {X.-W.}\ \bibnamefont {Guan}},\ }\bibfield  {title} {\bibinfo
  {title} {Exact quantum dynamics of {XXZ} central spin problems},\ }\href
  {https://doi.org/10.1103/PhysRevB.99.174308} {\bibfield  {journal} {\bibinfo
  {journal} {Phys. Rev. B}\ }\textbf {\bibinfo {volume} {99}},\ \bibinfo
  {pages} {174308} (\bibinfo {year} {2019})}\BibitemShut {NoStop}%
\bibitem [{\citenamefont {Schwartz}\ \emph {et~al.}(2018)\citenamefont
  {Schwartz}, \citenamefont {Scheuer}, \citenamefont {Tratzmiller},
  \citenamefont {M{\"u}ller}, \citenamefont {Chen}, \citenamefont {Dhand},
  \citenamefont {Wang}, \citenamefont {M{\"u}ller}, \citenamefont {Naydenov},
  \citenamefont {Jelezko} \emph {et~al.}}]{schwartz2018robust}%
  \BibitemOpen
  \bibfield  {author} {\bibinfo {author} {\bibfnamefont {I.}~\bibnamefont
  {Schwartz}}, \bibinfo {author} {\bibfnamefont {J.}~\bibnamefont {Scheuer}},
  \bibinfo {author} {\bibfnamefont {B.}~\bibnamefont {Tratzmiller}}, \bibinfo
  {author} {\bibfnamefont {S.}~\bibnamefont {M{\"u}ller}}, \bibinfo {author}
  {\bibfnamefont {Q.}~\bibnamefont {Chen}}, \bibinfo {author} {\bibfnamefont
  {I.}~\bibnamefont {Dhand}}, \bibinfo {author} {\bibfnamefont {Z.-Y.}\
  \bibnamefont {Wang}}, \bibinfo {author} {\bibfnamefont {C.}~\bibnamefont
  {M{\"u}ller}}, \bibinfo {author} {\bibfnamefont {B.}~\bibnamefont
  {Naydenov}}, \bibinfo {author} {\bibfnamefont {F.}~\bibnamefont {Jelezko}},
  \emph {et~al.},\ }\bibfield  {title} {\bibinfo {title} {Robust optical
  polarization of nuclear spin baths using {Hamiltonian} engineering of
  nitrogen-vacancy center quantum dynamics},\ }\href
  {https://doi.org/10.1126/sciadv.aat8978} {\bibfield  {journal} {\bibinfo
  {journal} {Sci. Adv.}\ }\textbf {\bibinfo {volume} {4}},\ \bibinfo {pages}
  {eaat8978} (\bibinfo {year} {2018})}\BibitemShut {NoStop}%
\bibitem [{\citenamefont {London}\ \emph {et~al.}(2013)\citenamefont {London},
  \citenamefont {Scheuer}, \citenamefont {Cai}, \citenamefont {Schwarz},
  \citenamefont {Retzker}, \citenamefont {Plenio}, \citenamefont {Katagiri},
  \citenamefont {Teraji}, \citenamefont {Koizumi}, \citenamefont {Isoya} \emph
  {et~al.}}]{london_polar_2013}%
  \BibitemOpen
  \bibfield  {author} {\bibinfo {author} {\bibfnamefont {P.}~\bibnamefont
  {London}}, \bibinfo {author} {\bibfnamefont {J.}~\bibnamefont {Scheuer}},
  \bibinfo {author} {\bibfnamefont {J.-M.}\ \bibnamefont {Cai}}, \bibinfo
  {author} {\bibfnamefont {I.}~\bibnamefont {Schwarz}}, \bibinfo {author}
  {\bibfnamefont {A.}~\bibnamefont {Retzker}}, \bibinfo {author} {\bibfnamefont
  {M.~B.}\ \bibnamefont {Plenio}}, \bibinfo {author} {\bibfnamefont
  {M.}~\bibnamefont {Katagiri}}, \bibinfo {author} {\bibfnamefont
  {T.}~\bibnamefont {Teraji}}, \bibinfo {author} {\bibfnamefont
  {S.}~\bibnamefont {Koizumi}}, \bibinfo {author} {\bibfnamefont
  {J.}~\bibnamefont {Isoya}}, \emph {et~al.},\ }\bibfield  {title} {\bibinfo
  {title} {Detecting and polarizing nuclear spins with double resonance on a
  single electron spin},\ }\href
  {https://doi.org/10.1103/PhysRevLett.111.067601} {\bibfield  {journal}
  {\bibinfo  {journal} {Phys. Rev. Lett.}\ }\textbf {\bibinfo {volume} {111}},\
  \bibinfo {pages} {067601} (\bibinfo {year} {2013})}\BibitemShut {NoStop}%
\bibitem [{\citenamefont {Hanson}\ \emph {et~al.}(2007)\citenamefont {Hanson},
  \citenamefont {Kouwenhoven}, \citenamefont {Petta}, \citenamefont {Tarucha},\
  and\ \citenamefont {Vandersypen}}]{hanson_review_2007}%
  \BibitemOpen
  \bibfield  {author} {\bibinfo {author} {\bibfnamefont {R.}~\bibnamefont
  {Hanson}}, \bibinfo {author} {\bibfnamefont {L.~P.}\ \bibnamefont
  {Kouwenhoven}}, \bibinfo {author} {\bibfnamefont {J.~R.}\ \bibnamefont
  {Petta}}, \bibinfo {author} {\bibfnamefont {S.}~\bibnamefont {Tarucha}},\
  and\ \bibinfo {author} {\bibfnamefont {L.~M.}\ \bibnamefont {Vandersypen}},\
  }\bibfield  {title} {\bibinfo {title} {Spins in few-electron quantum dots},\
  }\href {https://doi.org/10.1103/RevModPhys.79.1217} {\bibfield  {journal}
  {\bibinfo  {journal} {Rev. Mod. Phys.}\ }\textbf {\bibinfo {volume} {79}},\
  \bibinfo {pages} {1217} (\bibinfo {year} {2007})}\BibitemShut {NoStop}%
\bibitem [{\citenamefont {Schliemann}\ \emph {et~al.}(2003)\citenamefont
  {Schliemann}, \citenamefont {Khaetskii},\ and\ \citenamefont
  {Loss}}]{schliemann2003electron}%
  \BibitemOpen
  \bibfield  {author} {\bibinfo {author} {\bibfnamefont {J.}~\bibnamefont
  {Schliemann}}, \bibinfo {author} {\bibfnamefont {A.}~\bibnamefont
  {Khaetskii}},\ and\ \bibinfo {author} {\bibfnamefont {D.}~\bibnamefont
  {Loss}},\ }\bibfield  {title} {\bibinfo {title} {Electron spin dynamics in
  quantum dots and related nanostructures due to hyperfine interaction with
  nuclei},\ }\href {https://doi.org/10.1088/0953-8984/15/50/R01} {\bibfield
  {journal} {\bibinfo  {journal} {J. Phys. Condens. Matter}\ }\textbf {\bibinfo
  {volume} {15}},\ \bibinfo {pages} {R1809} (\bibinfo {year}
  {2003})}\BibitemShut {NoStop}%
\bibitem [{\citenamefont {Urbaszek}\ \emph {et~al.}(2013)\citenamefont
  {Urbaszek}, \citenamefont {Marie}, \citenamefont {Amand}, \citenamefont
  {Krebs}, \citenamefont {Voisin}, \citenamefont {Maletinsky}, \citenamefont
  {H{\"o}gele},\ and\ \citenamefont {Imamoglu}}]{urbaszek2013nuclear}%
  \BibitemOpen
  \bibfield  {author} {\bibinfo {author} {\bibfnamefont {B.}~\bibnamefont
  {Urbaszek}}, \bibinfo {author} {\bibfnamefont {X.}~\bibnamefont {Marie}},
  \bibinfo {author} {\bibfnamefont {T.}~\bibnamefont {Amand}}, \bibinfo
  {author} {\bibfnamefont {O.}~\bibnamefont {Krebs}}, \bibinfo {author}
  {\bibfnamefont {P.}~\bibnamefont {Voisin}}, \bibinfo {author} {\bibfnamefont
  {P.}~\bibnamefont {Maletinsky}}, \bibinfo {author} {\bibfnamefont
  {A.}~\bibnamefont {H{\"o}gele}},\ and\ \bibinfo {author} {\bibfnamefont
  {A.}~\bibnamefont {Imamoglu}},\ }\bibfield  {title} {\bibinfo {title}
  {Nuclear spin physics in quantum dots: An optical investigation},\ }\href
  {https://doi.org/https://doi.org/10.1103/RevModPhys.85.79} {\bibfield
  {journal} {\bibinfo  {journal} {Rev. Mod. Phys.}\ }\textbf {\bibinfo {volume}
  {85}},\ \bibinfo {pages} {79} (\bibinfo {year} {2013})}\BibitemShut {NoStop}%
\bibitem [{\citenamefont {Hartmann}\ and\ \citenamefont
  {Hahn}(1962)}]{hartmann1962nuclear}%
  \BibitemOpen
  \bibfield  {author} {\bibinfo {author} {\bibfnamefont {S.}~\bibnamefont
  {Hartmann}}\ and\ \bibinfo {author} {\bibfnamefont {E.}~\bibnamefont
  {Hahn}},\ }\bibfield  {title} {\bibinfo {title} {Nuclear double resonance in
  the rotating frame},\ }\href
  {https://doi.org/https://doi.org/10.1103/PhysRev.128.2042} {\bibfield
  {journal} {\bibinfo  {journal} {Phys. Rev.}\ }\textbf {\bibinfo {volume}
  {128}},\ \bibinfo {pages} {2042} (\bibinfo {year} {1962})}\BibitemShut
  {NoStop}%
\bibitem [{\citenamefont {Rovnyak}(2008)}]{rovnyak2008tutorial}%
  \BibitemOpen
  \bibfield  {author} {\bibinfo {author} {\bibfnamefont {D.}~\bibnamefont
  {Rovnyak}},\ }\bibfield  {title} {\bibinfo {title} {Tutorial on analytic
  theory for cross-polarization in solid state nmr},\ }\href
  {https://doi.org/10.1002/cmr.a.20115} {\bibfield  {journal} {\bibinfo
  {journal} {Conc. Magnet. Reson. A}\ }\textbf {\bibinfo {volume} {32}},\
  \bibinfo {pages} {254} (\bibinfo {year} {2008})}\BibitemShut {NoStop}%
\bibitem [{\citenamefont {Rao}\ \emph {et~al.}(2019)\citenamefont {Rao},
  \citenamefont {Ghosh}, \citenamefont {Gelbwaser-Klimovsky}, \citenamefont
  {Bar-Gill},\ and\ \citenamefont {Kurizki}}]{rao2019spin}%
  \BibitemOpen
  \bibfield  {author} {\bibinfo {author} {\bibfnamefont {D.}~\bibnamefont
  {Rao}}, \bibinfo {author} {\bibfnamefont {A.}~\bibnamefont {Ghosh}}, \bibinfo
  {author} {\bibfnamefont {D.}~\bibnamefont {Gelbwaser-Klimovsky}}, \bibinfo
  {author} {\bibfnamefont {N.}~\bibnamefont {Bar-Gill}},\ and\ \bibinfo
  {author} {\bibfnamefont {G.}~\bibnamefont {Kurizki}},\ }\bibfield  {title}
  {\bibinfo {title} {Spin-bath polarization via disentanglement},\ }\href
  {https://arxiv.org/abs/1912.00613} {\bibfield  {journal} {\bibinfo  {journal}
  {arXiv:1912.00613[quant-ph]}\ } (\bibinfo {year} {2019})}\BibitemShut
  {NoStop}%
\bibitem [{\citenamefont {Lai}\ \emph {et~al.}(2006)\citenamefont {Lai},
  \citenamefont {Maletinsky}, \citenamefont {Badolato},\ and\ \citenamefont
  {Imamoglu}}]{lai2006knight}%
  \BibitemOpen
  \bibfield  {author} {\bibinfo {author} {\bibfnamefont {C.}~\bibnamefont
  {Lai}}, \bibinfo {author} {\bibfnamefont {P.}~\bibnamefont {Maletinsky}},
  \bibinfo {author} {\bibfnamefont {A.}~\bibnamefont {Badolato}},\ and\
  \bibinfo {author} {\bibfnamefont {A.}~\bibnamefont {Imamoglu}},\ }\bibfield
  {title} {\bibinfo {title} {Knight-field-enabled nuclear spin polarization in
  single quantum dots},\ }\href
  {https://doi.org/https://doi.org/10.1103/PhysRevLett.96.167403} {\bibfield
  {journal} {\bibinfo  {journal} {Phys. Rev. Lett.}\ }\textbf {\bibinfo
  {volume} {96}},\ \bibinfo {pages} {167403} (\bibinfo {year}
  {2006})}\BibitemShut {NoStop}%
\bibitem [{\citenamefont {Ding}\ \emph {et~al.}(2014)\citenamefont {Ding},
  \citenamefont {Shi}, \citenamefont {You},\ and\ \citenamefont
  {Zhang}}]{ding2014high}%
  \BibitemOpen
  \bibfield  {author} {\bibinfo {author} {\bibfnamefont {W.}~\bibnamefont
  {Ding}}, \bibinfo {author} {\bibfnamefont {A.}~\bibnamefont {Shi}}, \bibinfo
  {author} {\bibfnamefont {J.}~\bibnamefont {You}},\ and\ \bibinfo {author}
  {\bibfnamefont {W.}~\bibnamefont {Zhang}},\ }\bibfield  {title} {\bibinfo
  {title} {High-fidelity quantum memory utilizing inhomogeneous nuclear
  polarization in a quantum dot},\ }\href
  {https://doi.org/https://doi.org/10.1103/PhysRevB.90.235421} {\bibfield
  {journal} {\bibinfo  {journal} {Phys. Rev. B}\ }\textbf {\bibinfo {volume}
  {90}},\ \bibinfo {pages} {235421} (\bibinfo {year} {2014})}\BibitemShut
  {NoStop}%
\bibitem [{\citenamefont {Taylor}\ \emph
  {et~al.}(2003{\natexlab{a}})\citenamefont {Taylor}, \citenamefont {Marcus},\
  and\ \citenamefont {Lukin}}]{taylor2003long}%
  \BibitemOpen
  \bibfield  {author} {\bibinfo {author} {\bibfnamefont {J.}~\bibnamefont
  {Taylor}}, \bibinfo {author} {\bibfnamefont {C.}~\bibnamefont {Marcus}},\
  and\ \bibinfo {author} {\bibfnamefont {M.}~\bibnamefont {Lukin}},\ }\bibfield
   {title} {\bibinfo {title} {Long-lived memory for mesoscopic quantum bits},\
  }\href {https://doi.org/https://doi.org/10.1103/PhysRevLett.90.206803}
  {\bibfield  {journal} {\bibinfo  {journal} {Phys. Rev. Lett.}\ }\textbf
  {\bibinfo {volume} {90}},\ \bibinfo {pages} {206803} (\bibinfo {year}
  {2003}{\natexlab{a}})}\BibitemShut {NoStop}%
\bibitem [{\citenamefont {Fern{\'a}ndez-Acebal}\ \emph
  {et~al.}(2018)\citenamefont {Fern{\'a}ndez-Acebal}, \citenamefont {Rosolio},
  \citenamefont {Scheuer}, \citenamefont {Müller}, \citenamefont {Müller},
  \citenamefont {Schmitt}, \citenamefont {McGuinness}, \citenamefont {Schwarz},
  \citenamefont {Chen}, \citenamefont {Retzker} \emph
  {et~al.}}]{fernandez-acebal_hyper_2017}%
  \BibitemOpen
  \bibfield  {author} {\bibinfo {author} {\bibfnamefont {P.}~\bibnamefont
  {Fern{\'a}ndez-Acebal}}, \bibinfo {author} {\bibfnamefont {O.}~\bibnamefont
  {Rosolio}}, \bibinfo {author} {\bibfnamefont {J.}~\bibnamefont {Scheuer}},
  \bibinfo {author} {\bibfnamefont {C.}~\bibnamefont {Müller}}, \bibinfo
  {author} {\bibfnamefont {S.}~\bibnamefont {Müller}}, \bibinfo {author}
  {\bibfnamefont {S.}~\bibnamefont {Schmitt}}, \bibinfo {author} {\bibfnamefont
  {L.~P.}\ \bibnamefont {McGuinness}}, \bibinfo {author} {\bibfnamefont
  {I.}~\bibnamefont {Schwarz}}, \bibinfo {author} {\bibfnamefont
  {Q.}~\bibnamefont {Chen}}, \bibinfo {author} {\bibfnamefont {A.}~\bibnamefont
  {Retzker}}, \emph {et~al.},\ }\bibfield  {title} {\bibinfo {title} {Toward
  hyperpolarization of oil molecules via single nitrogen vacancy centers in
  diamond},\ }\href {https://doi.org/10.1021/acs.nanolett.7b05175} {\bibfield
  {journal} {\bibinfo  {journal} {Nano Lett.}\ }\textbf {\bibinfo {volume}
  {18}},\ \bibinfo {pages} {1882} (\bibinfo {year} {2018})}\BibitemShut
  {NoStop}%
\bibitem [{\citenamefont {Gaudin}(2014)}]{gaudin_bethe_2014}%
  \BibitemOpen
  \bibfield  {author} {\bibinfo {author} {\bibfnamefont {M.}~\bibnamefont
  {Gaudin}},\ }\href {https://doi.org/10.1017/CBO9781107053885} {\emph
  {\bibinfo {title} {{T}he {Bethe} {Wavefunction}}}}\ (\bibinfo  {publisher}
  {Cambridge University Press},\ \bibinfo {address} {Cambridge},\ \bibinfo
  {year} {2014})\ \bibinfo {note} {{Translated by J.-S. Caux}}\BibitemShut
  {NoStop}%
\bibitem [{\citenamefont {Dukelsky}\ \emph {et~al.}(2004)\citenamefont
  {Dukelsky}, \citenamefont {Pittel},\ and\ \citenamefont
  {Sierra}}]{dukelsky_colloquium_2004}%
  \BibitemOpen
  \bibfield  {author} {\bibinfo {author} {\bibfnamefont {J.}~\bibnamefont
  {Dukelsky}}, \bibinfo {author} {\bibfnamefont {S.}~\bibnamefont {Pittel}},\
  and\ \bibinfo {author} {\bibfnamefont {G.}~\bibnamefont {Sierra}},\
  }\bibfield  {title} {\bibinfo {title} {{C}olloquium: {Exactly} solvable
  {Richardson}-{Gaudin} models for many-body quantum systems},\ }\href
  {https://doi.org/10.1103/RevModPhys.76.643} {\bibfield  {journal} {\bibinfo
  {journal} {Rev. Mod. Phys.}\ }\textbf {\bibinfo {volume} {76}},\ \bibinfo
  {pages} {643} (\bibinfo {year} {2004})}\BibitemShut {NoStop}%
\bibitem [{\citenamefont {Rombouts}\ \emph {et~al.}(2010)\citenamefont
  {Rombouts}, \citenamefont {Dukelsky},\ and\ \citenamefont
  {Ortiz}}]{rombouts_quantum_2010}%
  \BibitemOpen
  \bibfield  {author} {\bibinfo {author} {\bibfnamefont {S.~M.~A.}\
  \bibnamefont {Rombouts}}, \bibinfo {author} {\bibfnamefont {J.}~\bibnamefont
  {Dukelsky}},\ and\ \bibinfo {author} {\bibfnamefont {G.}~\bibnamefont
  {Ortiz}},\ }\bibfield  {title} {\bibinfo {title} {{Q}uantum phase diagram of
  the integrable $p_x+ip_y$ fermionic superfluid},\ }\href
  {https://doi.org/10.1103/PhysRevB.82.224510} {\bibfield  {journal} {\bibinfo
  {journal} {Phys. Rev. B}\ }\textbf {\bibinfo {volume} {82}},\ \bibinfo
  {pages} {224510} (\bibinfo {year} {2010})}\BibitemShut {NoStop}%
\bibitem [{\citenamefont {Bortz}\ and\ \citenamefont
  {Stolze}(2007)}]{bortz_exact_2007}%
  \BibitemOpen
  \bibfield  {author} {\bibinfo {author} {\bibfnamefont {M.}~\bibnamefont
  {Bortz}}\ and\ \bibinfo {author} {\bibfnamefont {J.}~\bibnamefont {Stolze}},\
  }\bibfield  {title} {\bibinfo {title} {Exact dynamics in the inhomogeneous
  central-spin model},\ }\href {https://doi.org/10.1103/PhysRevB.76.014304}
  {\bibfield  {journal} {\bibinfo  {journal} {Phys. Rev. B}\ }\textbf {\bibinfo
  {volume} {76}},\ \bibinfo {pages} {014304} (\bibinfo {year}
  {2007})}\BibitemShut {NoStop}%
\bibitem [{\citenamefont {Faribault}\ \emph {et~al.}(2009)\citenamefont
  {Faribault}, \citenamefont {Calabrese},\ and\ \citenamefont
  {Caux}}]{faribault_quantum_2009}%
  \BibitemOpen
  \bibfield  {author} {\bibinfo {author} {\bibfnamefont {A.}~\bibnamefont
  {Faribault}}, \bibinfo {author} {\bibfnamefont {P.}~\bibnamefont
  {Calabrese}},\ and\ \bibinfo {author} {\bibfnamefont {J.-S.}\ \bibnamefont
  {Caux}},\ }\bibfield  {title} {\bibinfo {title} {Quantum quenches from
  integrability: The fermionic pairing model},\ }\href
  {https://doi.org/10.1088/1742-5468/2009/03/P03018} {\bibfield  {journal}
  {\bibinfo  {journal} {J. Stat. Mech.}\ }\textbf {\bibinfo {volume} {2009}},\
  \bibinfo {pages} {P03018} (\bibinfo {year} {2009})}\BibitemShut {NoStop}%
\bibitem [{\citenamefont {Bortz}\ \emph {et~al.}(2010)\citenamefont {Bortz},
  \citenamefont {Eggert}, \citenamefont {Schneider}, \citenamefont {Stübner},\
  and\ \citenamefont {Stolze}}]{bortz_dynamics_2010}%
  \BibitemOpen
  \bibfield  {author} {\bibinfo {author} {\bibfnamefont {M.}~\bibnamefont
  {Bortz}}, \bibinfo {author} {\bibfnamefont {S.}~\bibnamefont {Eggert}},
  \bibinfo {author} {\bibfnamefont {C.}~\bibnamefont {Schneider}}, \bibinfo
  {author} {\bibfnamefont {R.}~\bibnamefont {Stübner}},\ and\ \bibinfo
  {author} {\bibfnamefont {J.}~\bibnamefont {Stolze}},\ }\bibfield  {title}
  {\bibinfo {title} {Dynamics and decoherence in the central spin model using
  exact methods},\ }\href {https://doi.org/10.1103/PhysRevB.82.161308}
  {\bibfield  {journal} {\bibinfo  {journal} {Phys. Rev. B}\ }\textbf {\bibinfo
  {volume} {82}},\ \bibinfo {pages} {161308} (\bibinfo {year}
  {2010})}\BibitemShut {NoStop}%
\bibitem [{\citenamefont {Faribault}\ and\ \citenamefont
  {Schuricht}(2013)}]{faribault_integrability-based_2013}%
  \BibitemOpen
  \bibfield  {author} {\bibinfo {author} {\bibfnamefont {A.}~\bibnamefont
  {Faribault}}\ and\ \bibinfo {author} {\bibfnamefont {D.}~\bibnamefont
  {Schuricht}},\ }\bibfield  {title} {\bibinfo {title} {Integrability-{based}
  {analysis} of the {hyperfine}-{interaction}-{induced} {decoherence} in
  {quantum} {dots}},\ }\href {https://doi.org/10.1103/PhysRevLett.110.040405}
  {\bibfield  {journal} {\bibinfo  {journal} {Phys. Rev. Lett.}\ }\textbf
  {\bibinfo {volume} {110}},\ \bibinfo {pages} {040405} (\bibinfo {year}
  {2013})}\BibitemShut {NoStop}%
\bibitem [{\citenamefont {Claeys}\ \emph {et~al.}(2018)\citenamefont {Claeys},
  \citenamefont {De~Baerdemacker}, \citenamefont {El~Araby},\ and\
  \citenamefont {Caux}}]{claeys_spin_2017}%
  \BibitemOpen
  \bibfield  {author} {\bibinfo {author} {\bibfnamefont {P.~W.}\ \bibnamefont
  {Claeys}}, \bibinfo {author} {\bibfnamefont {S.}~\bibnamefont
  {De~Baerdemacker}}, \bibinfo {author} {\bibfnamefont {O.}~\bibnamefont
  {El~Araby}},\ and\ \bibinfo {author} {\bibfnamefont {J.-S.}\ \bibnamefont
  {Caux}},\ }\bibfield  {title} {\bibinfo {title} {Spin {polarization} through
  {Floquet} {resonances} in a {driven} {central} {spin} {model}},\ }\href
  {https://doi.org/10.1103/PhysRevLett.121.080401} {\bibfield  {journal}
  {\bibinfo  {journal} {Phys. Rev. Lett.}\ }\textbf {\bibinfo {volume} {121}},\
  \bibinfo {pages} {080401} (\bibinfo {year} {2018})}\BibitemShut {NoStop}%
\bibitem [{\citenamefont {Jivulescu}\ \emph
  {et~al.}(2009{\natexlab{a}})\citenamefont {Jivulescu}, \citenamefont
  {Ferraro}, \citenamefont {Napoli},\ and\ \citenamefont
  {Messina}}]{jivulescu_exact_2009}%
  \BibitemOpen
  \bibfield  {author} {\bibinfo {author} {\bibfnamefont {M.~A.}\ \bibnamefont
  {Jivulescu}}, \bibinfo {author} {\bibfnamefont {E.}~\bibnamefont {Ferraro}},
  \bibinfo {author} {\bibfnamefont {A.}~\bibnamefont {Napoli}},\ and\ \bibinfo
  {author} {\bibfnamefont {A.}~\bibnamefont {Messina}},\ }\bibfield  {title}
  {\bibinfo {title} {Exact dynamics of {XX} central spin models},\ }\href
  {https://doi.org/10.1088/0031-8949/2009/T135/014049} {\bibfield  {journal}
  {\bibinfo  {journal} {Phys. Scr.}\ }\textbf {\bibinfo {volume} {T135}},\
  \bibinfo {pages} {014049} (\bibinfo {year} {2009}{\natexlab{a}})}\BibitemShut
  {NoStop}%
\bibitem [{\citenamefont {Jivulescu}\ \emph
  {et~al.}(2009{\natexlab{b}})\citenamefont {Jivulescu}, \citenamefont
  {Ferraro}, \citenamefont {Napoli},\ and\ \citenamefont
  {Messina}}]{jivulescu_dynamical_2009}%
  \BibitemOpen
  \bibfield  {author} {\bibinfo {author} {\bibfnamefont {M.}~\bibnamefont
  {Jivulescu}}, \bibinfo {author} {\bibfnamefont {E.}~\bibnamefont {Ferraro}},
  \bibinfo {author} {\bibfnamefont {A.}~\bibnamefont {Napoli}},\ and\ \bibinfo
  {author} {\bibfnamefont {A.}~\bibnamefont {Messina}},\ }\bibfield  {title}
  {\bibinfo {title} {Dynamical behaviour of an {XX} central spin model through
  {Bethe} ansatz techniques},\ }\href
  {https://doi.org/10.1016/S0034-4877(09)90036-2} {\bibfield  {journal}
  {\bibinfo  {journal} {Rep. Math. Phys.}\ }\textbf {\bibinfo {volume} {64}},\
  \bibinfo {pages} {315} (\bibinfo {year} {2009}{\natexlab{b}})}\BibitemShut
  {NoStop}%
\bibitem [{\citenamefont {Taylor}\ \emph
  {et~al.}(2003{\natexlab{b}})\citenamefont {Taylor}, \citenamefont
  {Imamoglu},\ and\ \citenamefont {Lukin}}]{taylor_controlling_2003}%
  \BibitemOpen
  \bibfield  {author} {\bibinfo {author} {\bibfnamefont {J.~M.}\ \bibnamefont
  {Taylor}}, \bibinfo {author} {\bibfnamefont {A.}~\bibnamefont {Imamoglu}},\
  and\ \bibinfo {author} {\bibfnamefont {M.~D.}\ \bibnamefont {Lukin}},\
  }\bibfield  {title} {\bibinfo {title} {Controlling a {mesoscopic} {spin}
  {environment} by {quantum} {bit} {manipulation}},\ }\href
  {https://doi.org/10.1103/PhysRevLett.91.246802} {\bibfield  {journal}
  {\bibinfo  {journal} {Phys. Rev. Lett.}\ }\textbf {\bibinfo {volume} {91}},\
  \bibinfo {pages} {246802} (\bibinfo {year} {2003}{\natexlab{b}})}\BibitemShut
  {NoStop}%
\bibitem [{\citenamefont {Imamoḡlu}\ \emph {et~al.}(2003)\citenamefont
  {Imamoḡlu}, \citenamefont {Knill}, \citenamefont {Tian},\ and\
  \citenamefont {Zoller}}]{imamoglu_optical_2003}%
  \BibitemOpen
  \bibfield  {author} {\bibinfo {author} {\bibfnamefont {A.}~\bibnamefont
  {Imamoḡlu}}, \bibinfo {author} {\bibfnamefont {E.}~\bibnamefont {Knill}},
  \bibinfo {author} {\bibfnamefont {L.}~\bibnamefont {Tian}},\ and\ \bibinfo
  {author} {\bibfnamefont {P.}~\bibnamefont {Zoller}},\ }\bibfield  {title}
  {\bibinfo {title} {Optical {pumping} of {quantum}-{dot} {nuclear} {spins}},\
  }\href {https://doi.org/10.1103/PhysRevLett.91.017402} {\bibfield  {journal}
  {\bibinfo  {journal} {Phys. Rev. Lett.}\ }\textbf {\bibinfo {volume} {91}},\
  \bibinfo {pages} {017402} (\bibinfo {year} {2003})}\BibitemShut {NoStop}%
\bibitem [{\citenamefont {Christ}\ \emph {et~al.}(2009)\citenamefont {Christ},
  \citenamefont {Cirac},\ and\ \citenamefont {Giedke}}]{christ_nuclear_2009}%
  \BibitemOpen
  \bibfield  {author} {\bibinfo {author} {\bibfnamefont {H.}~\bibnamefont
  {Christ}}, \bibinfo {author} {\bibfnamefont {J.}~\bibnamefont {Cirac}},\ and\
  \bibinfo {author} {\bibfnamefont {G.}~\bibnamefont {Giedke}},\ }\bibfield
  {title} {\bibinfo {title} {Nuclear spin polarization in quantum dots -- {The}
  homogeneous limit},\ }\href
  {https://doi.org/10.1016/j.solidstatesciences.2007.09.027} {\bibfield
  {journal} {\bibinfo  {journal} {Solid State Sci.}\ }\textbf {\bibinfo
  {volume} {11}},\ \bibinfo {pages} {965} (\bibinfo {year} {2009})}\BibitemShut
  {NoStop}%
\bibitem [{\citenamefont {Belthangady}\ \emph {et~al.}(2013)\citenamefont
  {Belthangady}, \citenamefont {Bar-Gill}, \citenamefont {Pham}, \citenamefont
  {Arai}, \citenamefont {Le~Sage}, \citenamefont {Cappellaro},\ and\
  \citenamefont {Walsworth}}]{belthangady2013dressed}%
  \BibitemOpen
  \bibfield  {author} {\bibinfo {author} {\bibfnamefont {C.}~\bibnamefont
  {Belthangady}}, \bibinfo {author} {\bibfnamefont {N.}~\bibnamefont
  {Bar-Gill}}, \bibinfo {author} {\bibfnamefont {L.~M.}\ \bibnamefont {Pham}},
  \bibinfo {author} {\bibfnamefont {K.}~\bibnamefont {Arai}}, \bibinfo {author}
  {\bibfnamefont {D.}~\bibnamefont {Le~Sage}}, \bibinfo {author} {\bibfnamefont
  {P.}~\bibnamefont {Cappellaro}},\ and\ \bibinfo {author} {\bibfnamefont
  {R.~L.}\ \bibnamefont {Walsworth}},\ }\bibfield  {title} {\bibinfo {title}
  {Dressed-state resonant coupling between bright and dark spins in diamond},\
  }\href {https://doi.org/10.1103/PhysRevLett.110.157601} {\bibfield  {journal}
  {\bibinfo  {journal} {Phys. Rev. Lett.}\ }\textbf {\bibinfo {volume} {110}},\
  \bibinfo {pages} {157601} (\bibinfo {year} {2013})}\BibitemShut {NoStop}%
\bibitem [{\citenamefont {Kurucz}\ \emph {et~al.}(2009)\citenamefont {Kurucz},
  \citenamefont {S{\o}rensen}, \citenamefont {Taylor}, \citenamefont {Lukin},\
  and\ \citenamefont {Fleischhauer}}]{kurucz2009qubit}%
  \BibitemOpen
  \bibfield  {author} {\bibinfo {author} {\bibfnamefont {Z.}~\bibnamefont
  {Kurucz}}, \bibinfo {author} {\bibfnamefont {M.~W.}\ \bibnamefont
  {S{\o}rensen}}, \bibinfo {author} {\bibfnamefont {J.~M.}\ \bibnamefont
  {Taylor}}, \bibinfo {author} {\bibfnamefont {M.~D.}\ \bibnamefont {Lukin}},\
  and\ \bibinfo {author} {\bibfnamefont {M.}~\bibnamefont {Fleischhauer}},\
  }\bibfield  {title} {\bibinfo {title} {Qubit protection in nuclear-spin
  quantum dot memories},\ }\href
  {https://doi.org/https://doi.org/10.1103/PhysRevLett.103.010502} {\bibfield
  {journal} {\bibinfo  {journal} {Phys. Rev. Lett.}\ }\textbf {\bibinfo
  {volume} {103}},\ \bibinfo {pages} {010502} (\bibinfo {year}
  {2009})}\BibitemShut {NoStop}%
\bibitem [{\citenamefont {Claeys}\ \emph {et~al.}(2017)\citenamefont {Claeys},
  \citenamefont {De~Baerdemacker},\ and\ \citenamefont
  {Van~Neck}}]{claeys_inner_2017}%
  \BibitemOpen
  \bibfield  {author} {\bibinfo {author} {\bibfnamefont {P.~W.}\ \bibnamefont
  {Claeys}}, \bibinfo {author} {\bibfnamefont {S.}~\bibnamefont
  {De~Baerdemacker}},\ and\ \bibinfo {author} {\bibfnamefont {D.}~\bibnamefont
  {Van~Neck}},\ }\bibfield  {title} {\bibinfo {title} {Inner products in
  integrable {Richardson}-{Gaudin} models},\ }\href
  {https://doi.org/10.21468/SciPostPhys.3.4.028} {\bibfield  {journal}
  {\bibinfo  {journal} {SciPost Phys.}\ }\textbf {\bibinfo {volume} {3}},\
  \bibinfo {pages} {028} (\bibinfo {year} {2017})}\BibitemShut {NoStop}%
\bibitem [{\citenamefont {Claeys}\ \emph {et~al.}(2015)\citenamefont {Claeys},
  \citenamefont {De~Baerdemacker}, \citenamefont {Van~Raemdonck},\ and\
  \citenamefont {Van~Neck}}]{claeys_eigenvalue-based_2015}%
  \BibitemOpen
  \bibfield  {author} {\bibinfo {author} {\bibfnamefont {P.~W.}\ \bibnamefont
  {Claeys}}, \bibinfo {author} {\bibfnamefont {S.}~\bibnamefont
  {De~Baerdemacker}}, \bibinfo {author} {\bibfnamefont {M.}~\bibnamefont
  {Van~Raemdonck}},\ and\ \bibinfo {author} {\bibfnamefont {D.}~\bibnamefont
  {Van~Neck}},\ }\bibfield  {title} {\bibinfo {title} {{E}igenvalue-based
  method and form-factor determinant representations for integrable {XXZ}
  {Richardson}-{Gaudin} models},\ }\href
  {https://doi.org/10.1103/PhysRevB.91.155102} {\bibfield  {journal} {\bibinfo
  {journal} {Phys. Rev. B}\ }\textbf {\bibinfo {volume} {91}},\ \bibinfo
  {pages} {155102} (\bibinfo {year} {2015})}\BibitemShut {NoStop}%
\bibitem [{\citenamefont {Faribault}\ \emph {et~al.}(2011)\citenamefont
  {Faribault}, \citenamefont {El~Araby}, \citenamefont {Str{\"a}ter},\ and\
  \citenamefont {Gritsev}}]{faribault_gaudin_2011}%
  \BibitemOpen
  \bibfield  {author} {\bibinfo {author} {\bibfnamefont {A.}~\bibnamefont
  {Faribault}}, \bibinfo {author} {\bibfnamefont {O.}~\bibnamefont {El~Araby}},
  \bibinfo {author} {\bibfnamefont {C.}~\bibnamefont {Str{\"a}ter}},\ and\
  \bibinfo {author} {\bibfnamefont {V.}~\bibnamefont {Gritsev}},\ }\bibfield
  {title} {\bibinfo {title} {{G}audin models solver based on the correspondence
  between {Bethe} ansatz and ordinary differential equations},\ }\href
  {https://doi.org/10.1103/PhysRevB.83.235124} {\bibfield  {journal} {\bibinfo
  {journal} {Phys. Rev. B}\ }\textbf {\bibinfo {volume} {83}},\ \bibinfo
  {pages} {235124} (\bibinfo {year} {2011})}\BibitemShut {NoStop}%
\bibitem [{\citenamefont {Links}(2017)}]{links_completeness_2016}%
  \BibitemOpen
  \bibfield  {author} {\bibinfo {author} {\bibfnamefont {J.}~\bibnamefont
  {Links}},\ }\bibfield  {title} {\bibinfo {title} {Completeness of the {Bethe}
  states for the rational, spin-1/2 {Richardson}-{Gaudin} system},\ }\href
  {https://doi.org/10.21468/SciPostPhys.3.1.007} {\bibfield  {journal}
  {\bibinfo  {journal} {SciPost Phys.}\ }\textbf {\bibinfo {volume} {3}},\
  \bibinfo {pages} {007} (\bibinfo {year} {2017})}\BibitemShut {NoStop}%
\bibitem [{Note1()}]{Note1}%
  \BibitemOpen
  \bibinfo {note} {Any eigenstate can be constructed in two ways, either by
  expressing the state in terms of generalized spin raising operators acting on
  a vacuum state with all spins maximally down, or in terms of spin lowering
  operators acting on a dual vacuum state with all spins maximally up. For
  $S^z<0$, the former is more transparent and highlights the differences
  between bright states and dark states (with central spin necessarily down),
  and for $S^z>0$ the latter is more convenient since all dark states then have
  central spin up.}\BibitemShut {Stop}%
\bibitem [{\citenamefont {Essler}\ \emph {et~al.}(2005)\citenamefont {Essler},
  \citenamefont {Frahm}, \citenamefont {Göhmann}, \citenamefont {Klümper},\
  and\ \citenamefont {Korepin}}]{essler_one-dimensional_2005}%
  \BibitemOpen
  \bibfield  {author} {\bibinfo {author} {\bibfnamefont {F.~H.}\ \bibnamefont
  {Essler}}, \bibinfo {author} {\bibfnamefont {H.}~\bibnamefont {Frahm}},
  \bibinfo {author} {\bibfnamefont {F.}~\bibnamefont {Göhmann}}, \bibinfo
  {author} {\bibfnamefont {A.}~\bibnamefont {Klümper}},\ and\ \bibinfo
  {author} {\bibfnamefont {V.~E.}\ \bibnamefont {Korepin}},\ }\href
  {http://bilder.buecher.de/zusatz/14/14956/14956838_vorw_1.pdf} {\emph
  {\bibinfo {title} {The one-dimensional {Hubbard} model}}},\ Vol.\ \bibinfo
  {volume} {690}\ (\bibinfo  {publisher} {Cambridge University Press
  Cambridge},\ \bibinfo {year} {2005})\BibitemShut {NoStop}%
\bibitem [{\citenamefont {Iba{\~n}ez}\ \emph {et~al.}(2009)\citenamefont
  {Iba{\~n}ez}, \citenamefont {Links}, \citenamefont {Sierra},\ and\
  \citenamefont {Zhao}}]{ibanez_exactly_2009}%
  \BibitemOpen
  \bibfield  {author} {\bibinfo {author} {\bibfnamefont {M.}~\bibnamefont
  {Iba{\~n}ez}}, \bibinfo {author} {\bibfnamefont {J.}~\bibnamefont {Links}},
  \bibinfo {author} {\bibfnamefont {G.}~\bibnamefont {Sierra}},\ and\ \bibinfo
  {author} {\bibfnamefont {S.-Y.}\ \bibnamefont {Zhao}},\ }\bibfield  {title}
  {\bibinfo {title} {{E}xactly solvable pairing model for superconductors with
  $p_x+ip_y$-wave symmetry},\ }\href
  {https://doi.org/10.1103/PhysRevB.79.180501} {\bibfield  {journal} {\bibinfo
  {journal} {Phys. Rev. B}\ }\textbf {\bibinfo {volume} {79}},\ \bibinfo
  {pages} {180501} (\bibinfo {year} {2009})}\BibitemShut {NoStop}%
\bibitem [{\citenamefont {Lerma~H.}\ \emph {et~al.}(2011)\citenamefont
  {Lerma~H.}, \citenamefont {Rombouts}, \citenamefont {Dukelsky},\ and\
  \citenamefont {Ortiz}}]{lerma_h._integrable_2011}%
  \BibitemOpen
  \bibfield  {author} {\bibinfo {author} {\bibfnamefont {S.}~\bibnamefont
  {Lerma~H.}}, \bibinfo {author} {\bibfnamefont {S.~M.~A.}\ \bibnamefont
  {Rombouts}}, \bibinfo {author} {\bibfnamefont {J.}~\bibnamefont {Dukelsky}},\
  and\ \bibinfo {author} {\bibfnamefont {G.}~\bibnamefont {Ortiz}},\ }\bibfield
   {title} {\bibinfo {title} {Integrable two-channel $p_x+ip_y$-wave model of a
  superfluid},\ }\href {https://doi.org/10.1103/PhysRevB.84.100503} {\bibfield
  {journal} {\bibinfo  {journal} {Phys. Rev. B}\ }\textbf {\bibinfo {volume}
  {84}},\ \bibinfo {pages} {100503} (\bibinfo {year} {2011})}\BibitemShut
  {NoStop}%
\bibitem [{\citenamefont {Ortiz}\ \emph {et~al.}(2014)\citenamefont {Ortiz},
  \citenamefont {Dukelsky}, \citenamefont {Cobanera}, \citenamefont {Esebbag},\
  and\ \citenamefont {Beenakker}}]{ortiz_many-body_2014}%
  \BibitemOpen
  \bibfield  {author} {\bibinfo {author} {\bibfnamefont {G.}~\bibnamefont
  {Ortiz}}, \bibinfo {author} {\bibfnamefont {J.}~\bibnamefont {Dukelsky}},
  \bibinfo {author} {\bibfnamefont {E.}~\bibnamefont {Cobanera}}, \bibinfo
  {author} {\bibfnamefont {C.}~\bibnamefont {Esebbag}},\ and\ \bibinfo {author}
  {\bibfnamefont {C.}~\bibnamefont {Beenakker}},\ }\bibfield  {title} {\bibinfo
  {title} {Many-{body} {characterization} of {particle}-{conserving}
  {topological} {superfluids}},\ }\href
  {https://doi.org/10.1103/PhysRevLett.113.267002} {\bibfield  {journal}
  {\bibinfo  {journal} {Phys. Rev. Lett.}\ }\textbf {\bibinfo {volume} {113}},\
  \bibinfo {pages} {267002} (\bibinfo {year} {2014})}\BibitemShut {NoStop}%
\bibitem [{\citenamefont {Dean}\ and\ \citenamefont
  {Hjorth-Jensen}(2003)}]{dean_pairing_2003}%
  \BibitemOpen
  \bibfield  {author} {\bibinfo {author} {\bibfnamefont {D.~J.}\ \bibnamefont
  {Dean}}\ and\ \bibinfo {author} {\bibfnamefont {M.}~\bibnamefont
  {Hjorth-Jensen}},\ }\bibfield  {title} {\bibinfo {title} {Pairing in nuclear
  systems: from neutron stars to finite nuclei},\ }\href
  {https://doi.org/10.1103/RevModPhys.75.607} {\bibfield  {journal} {\bibinfo
  {journal} {Rev. Mod. Phys.}\ }\textbf {\bibinfo {volume} {75}},\ \bibinfo
  {pages} {607} (\bibinfo {year} {2003})}\BibitemShut {NoStop}%
\bibitem [{\citenamefont {Iyoda}\ \emph {et~al.}(2018)\citenamefont {Iyoda},
  \citenamefont {Katsura},\ and\ \citenamefont
  {Sagawa}}]{iyoda_effective_2018}%
  \BibitemOpen
  \bibfield  {author} {\bibinfo {author} {\bibfnamefont {E.}~\bibnamefont
  {Iyoda}}, \bibinfo {author} {\bibfnamefont {H.}~\bibnamefont {Katsura}},\
  and\ \bibinfo {author} {\bibfnamefont {T.}~\bibnamefont {Sagawa}},\
  }\bibfield  {title} {\bibinfo {title} {Effective dimension, level statistics,
  and integrability of {Sachdev}-{Ye}-{Kitaev}-like models},\ }\href
  {https://doi.org/10.1103/PhysRevD.98.086020} {\bibfield  {journal} {\bibinfo
  {journal} {Phys. Rev. D}\ }\textbf {\bibinfo {volume} {98}},\ \bibinfo
  {pages} {086020} (\bibinfo {year} {2018})}\BibitemShut {NoStop}%
\bibitem [{\citenamefont {Dukelsky}\ \emph {et~al.}(2001)\citenamefont
  {Dukelsky}, \citenamefont {Esebbag},\ and\ \citenamefont
  {Schuck}}]{dukelsky_class_2001}%
  \BibitemOpen
  \bibfield  {author} {\bibinfo {author} {\bibfnamefont {J.}~\bibnamefont
  {Dukelsky}}, \bibinfo {author} {\bibfnamefont {C.}~\bibnamefont {Esebbag}},\
  and\ \bibinfo {author} {\bibfnamefont {P.}~\bibnamefont {Schuck}},\
  }\bibfield  {title} {\bibinfo {title} {{C}lass of exactly solvable pairing
  models},\ }\href {https://doi.org/10.1103/PhysRevLett.87.066403} {\bibfield
  {journal} {\bibinfo  {journal} {Phys. Rev. Lett.}\ }\textbf {\bibinfo
  {volume} {87}},\ \bibinfo {pages} {066403} (\bibinfo {year}
  {2001})}\BibitemShut {NoStop}%
\bibitem [{\citenamefont {Ortiz}\ \emph {et~al.}(2005)\citenamefont {Ortiz},
  \citenamefont {Somma}, \citenamefont {Dukelsky},\ and\ \citenamefont
  {Rombouts}}]{ortiz_exactly-solvable_2005}%
  \BibitemOpen
  \bibfield  {author} {\bibinfo {author} {\bibfnamefont {G.}~\bibnamefont
  {Ortiz}}, \bibinfo {author} {\bibfnamefont {R.}~\bibnamefont {Somma}},
  \bibinfo {author} {\bibfnamefont {J.}~\bibnamefont {Dukelsky}},\ and\
  \bibinfo {author} {\bibfnamefont {S.}~\bibnamefont {Rombouts}},\ }\bibfield
  {title} {\bibinfo {title} {{E}xactly-solvable models derived from a
  generalized {Gaudin} algebra},\ }\href
  {https://doi.org/10.1016/j.nuclphysb.2004.11.008} {\bibfield  {journal}
  {\bibinfo  {journal} {Nucl. Phys. B}\ }\textbf {\bibinfo {volume} {707}},\
  \bibinfo {pages} {421} (\bibinfo {year} {2005})}\BibitemShut {NoStop}%
\bibitem [{\citenamefont {Van~Raemdonck}\ \emph {et~al.}(2014)\citenamefont
  {Van~Raemdonck}, \citenamefont {De~Baerdemacker},\ and\ \citenamefont
  {Van~Neck}}]{van_raemdonck_exact_2014}%
  \BibitemOpen
  \bibfield  {author} {\bibinfo {author} {\bibfnamefont {M.}~\bibnamefont
  {Van~Raemdonck}}, \bibinfo {author} {\bibfnamefont {S.}~\bibnamefont
  {De~Baerdemacker}},\ and\ \bibinfo {author} {\bibfnamefont {D.}~\bibnamefont
  {Van~Neck}},\ }\bibfield  {title} {\bibinfo {title} {{E}xact solution of the
  $p_x+ip_y$ pairing {Hamiltonian} by deforming the pairing algebra},\ }\href
  {https://doi.org/10.1103/PhysRevB.89.155136} {\bibfield  {journal} {\bibinfo
  {journal} {Phys. Rev. B}\ }\textbf {\bibinfo {volume} {89}},\ \bibinfo
  {pages} {155136} (\bibinfo {year} {2014})}\BibitemShut {NoStop}%
\bibitem [{\citenamefont {Links}\ \emph {et~al.}(2015)\citenamefont {Links},
  \citenamefont {Marquette},\ and\ \citenamefont
  {Moghaddam}}]{links_exact_2015}%
  \BibitemOpen
  \bibfield  {author} {\bibinfo {author} {\bibfnamefont {J.}~\bibnamefont
  {Links}}, \bibinfo {author} {\bibfnamefont {I.}~\bibnamefont {Marquette}},\
  and\ \bibinfo {author} {\bibfnamefont {A.}~\bibnamefont {Moghaddam}},\
  }\bibfield  {title} {\bibinfo {title} {{E}xact solution of the $p+ip$
  {Hamiltonian} revisited: duality relations in the hole-pair picture},\ }\href
  {https://doi.org/10.1088/1751-8113/48/37/374001} {\bibfield  {journal}
  {\bibinfo  {journal} {J. Phys. A: Math. Theor.}\ }\textbf {\bibinfo {volume}
  {48}},\ \bibinfo {pages} {374001} (\bibinfo {year} {2015})}\BibitemShut
  {NoStop}%
\bibitem [{\citenamefont {Bracker}\ \emph {et~al.}(2005)\citenamefont
  {Bracker}, \citenamefont {Stinaff}, \citenamefont {Gammon}, \citenamefont
  {Ware}, \citenamefont {Tischler}, \citenamefont {Shabaev}, \citenamefont
  {Efros}, \citenamefont {Park}, \citenamefont {Gershoni}, \citenamefont
  {Korenev} \emph {et~al.}}]{bracker2005optical}%
  \BibitemOpen
  \bibfield  {author} {\bibinfo {author} {\bibfnamefont {A.}~\bibnamefont
  {Bracker}}, \bibinfo {author} {\bibfnamefont {E.}~\bibnamefont {Stinaff}},
  \bibinfo {author} {\bibfnamefont {D.}~\bibnamefont {Gammon}}, \bibinfo
  {author} {\bibfnamefont {M.}~\bibnamefont {Ware}}, \bibinfo {author}
  {\bibfnamefont {J.}~\bibnamefont {Tischler}}, \bibinfo {author}
  {\bibfnamefont {A.}~\bibnamefont {Shabaev}}, \bibinfo {author} {\bibfnamefont
  {A.~L.}\ \bibnamefont {Efros}}, \bibinfo {author} {\bibfnamefont
  {D.}~\bibnamefont {Park}}, \bibinfo {author} {\bibfnamefont {D.}~\bibnamefont
  {Gershoni}}, \bibinfo {author} {\bibfnamefont {V.}~\bibnamefont {Korenev}},
  \emph {et~al.},\ }\bibfield  {title} {\bibinfo {title} {Optical pumping of
  the electronic and nuclear spin of single charge-tunable quantum dots},\
  }\href {https://doi.org/https://doi.org/10.1103/PhysRevLett.94.047402}
  {\bibfield  {journal} {\bibinfo  {journal} {Phys. Rev. Lett.}\ }\textbf
  {\bibinfo {volume} {94}},\ \bibinfo {pages} {047402} (\bibinfo {year}
  {2005})}\BibitemShut {NoStop}%
\bibitem [{\citenamefont {Urbaszek}\ \emph {et~al.}(2007)\citenamefont
  {Urbaszek}, \citenamefont {Braun}, \citenamefont {Amand}, \citenamefont
  {Krebs}, \citenamefont {Belhadj}, \citenamefont {Lema{\'\i}tre},
  \citenamefont {Voisin},\ and\ \citenamefont {Marie}}]{urbaszek2007efficient}%
  \BibitemOpen
  \bibfield  {author} {\bibinfo {author} {\bibfnamefont {B.}~\bibnamefont
  {Urbaszek}}, \bibinfo {author} {\bibfnamefont {P.-F.}\ \bibnamefont {Braun}},
  \bibinfo {author} {\bibfnamefont {T.}~\bibnamefont {Amand}}, \bibinfo
  {author} {\bibfnamefont {O.}~\bibnamefont {Krebs}}, \bibinfo {author}
  {\bibfnamefont {T.}~\bibnamefont {Belhadj}}, \bibinfo {author} {\bibfnamefont
  {A.}~\bibnamefont {Lema{\'\i}tre}}, \bibinfo {author} {\bibfnamefont
  {P.}~\bibnamefont {Voisin}},\ and\ \bibinfo {author} {\bibfnamefont
  {X.}~\bibnamefont {Marie}},\ }\bibfield  {title} {\bibinfo {title} {Efficient
  dynamical nuclear polarization in quantum dots: Temperature dependence},\
  }\href {https://doi.org/https://doi.org/10.1103/PhysRevB.76.201301}
  {\bibfield  {journal} {\bibinfo  {journal} {Phys. Rev. B}\ }\textbf {\bibinfo
  {volume} {76}},\ \bibinfo {pages} {201301} (\bibinfo {year}
  {2007})}\BibitemShut {NoStop}%
\bibitem [{\citenamefont {Chekhovich}\ \emph {et~al.}(2010)\citenamefont
  {Chekhovich}, \citenamefont {Makhonin}, \citenamefont {Kavokin},
  \citenamefont {Krysa}, \citenamefont {Skolnick},\ and\ \citenamefont
  {Tartakovskii}}]{chekhovich2010pumping}%
  \BibitemOpen
  \bibfield  {author} {\bibinfo {author} {\bibfnamefont {E.}~\bibnamefont
  {Chekhovich}}, \bibinfo {author} {\bibfnamefont {M.}~\bibnamefont
  {Makhonin}}, \bibinfo {author} {\bibfnamefont {K.}~\bibnamefont {Kavokin}},
  \bibinfo {author} {\bibfnamefont {A.}~\bibnamefont {Krysa}}, \bibinfo
  {author} {\bibfnamefont {M.}~\bibnamefont {Skolnick}},\ and\ \bibinfo
  {author} {\bibfnamefont {A.}~\bibnamefont {Tartakovskii}},\ }\bibfield
  {title} {\bibinfo {title} {Pumping of nuclear spins by optical excitation of
  spin-forbidden transitions in a quantum dot},\ }\href
  {https://doi.org/https://doi.org/10.1103/PhysRevLett.104.066804} {\bibfield
  {journal} {\bibinfo  {journal} {Phys. Rev. Lett.}\ }\textbf {\bibinfo
  {volume} {104}},\ \bibinfo {pages} {066804} (\bibinfo {year}
  {2010})}\BibitemShut {NoStop}%
\bibitem [{\citenamefont {Jaynes}(1957)}]{jaynes1957information}%
  \BibitemOpen
  \bibfield  {author} {\bibinfo {author} {\bibfnamefont {E.~T.}\ \bibnamefont
  {Jaynes}},\ }\bibfield  {title} {\bibinfo {title} {Information theory and
  statistical mechanics},\ }\href
  {https://doi.org/https://doi.org/10.1103/PhysRev.106.620} {\bibfield
  {journal} {\bibinfo  {journal} {Phys. Rev.}\ }\textbf {\bibinfo {volume}
  {106}},\ \bibinfo {pages} {620} (\bibinfo {year} {1957})}\BibitemShut
  {NoStop}%
\bibitem [{\citenamefont {Rigol}\ \emph {et~al.}(2007)\citenamefont {Rigol},
  \citenamefont {Dunjko}, \citenamefont {Yurovsky},\ and\ \citenamefont
  {Olshanii}}]{rigol2007relaxation}%
  \BibitemOpen
  \bibfield  {author} {\bibinfo {author} {\bibfnamefont {M.}~\bibnamefont
  {Rigol}}, \bibinfo {author} {\bibfnamefont {V.}~\bibnamefont {Dunjko}},
  \bibinfo {author} {\bibfnamefont {V.}~\bibnamefont {Yurovsky}},\ and\
  \bibinfo {author} {\bibfnamefont {M.}~\bibnamefont {Olshanii}},\ }\bibfield
  {title} {\bibinfo {title} {Relaxation in a completely integrable many-body
  quantum system: {An} ab initio study of the dynamics of the highly excited
  states of 1d lattice hard-core bosons},\ }\href
  {https://doi.org/https://doi.org/10.1103/PhysRevLett.98.050405} {\bibfield
  {journal} {\bibinfo  {journal} {Phys. Rev. Lett.}\ }\textbf {\bibinfo
  {volume} {98}},\ \bibinfo {pages} {050405} (\bibinfo {year}
  {2007})}\BibitemShut {NoStop}%
\bibitem [{\citenamefont {Vidmar}\ and\ \citenamefont
  {Rigol}(2016)}]{vidmar2016generalized}%
  \BibitemOpen
  \bibfield  {author} {\bibinfo {author} {\bibfnamefont {L.}~\bibnamefont
  {Vidmar}}\ and\ \bibinfo {author} {\bibfnamefont {M.}~\bibnamefont {Rigol}},\
  }\bibfield  {title} {\bibinfo {title} {Generalized {Gibbs} ensemble in
  integrable lattice models},\ }\href
  {https://iopscience.iop.org/article/10.1088/1742-5468/2016/06/064007/meta}
  {\bibfield  {journal} {\bibinfo  {journal} {J. Stat. Mech.}\ }\textbf
  {\bibinfo {volume} {2016}},\ \bibinfo {pages} {064007} (\bibinfo {year}
  {2016})}\BibitemShut {NoStop}%
\bibitem [{\citenamefont {Faribault}\ \emph {et~al.}(2008)\citenamefont
  {Faribault}, \citenamefont {Calabrese},\ and\ \citenamefont
  {Caux}}]{faribault_exact_2008}%
  \BibitemOpen
  \bibfield  {author} {\bibinfo {author} {\bibfnamefont {A.}~\bibnamefont
  {Faribault}}, \bibinfo {author} {\bibfnamefont {P.}~\bibnamefont
  {Calabrese}},\ and\ \bibinfo {author} {\bibfnamefont {J.-S.}\ \bibnamefont
  {Caux}},\ }\bibfield  {title} {\bibinfo {title} {{E}xact mesoscopic
  correlation functions of the {Richardson} pairing model},\ }\href
  {https://doi.org/10.1103/PhysRevB.77.064503} {\bibfield  {journal} {\bibinfo
  {journal} {Phys. Rev. B}\ }\textbf {\bibinfo {volume} {77}},\ \bibinfo
  {pages} {064503} (\bibinfo {year} {2008})}\BibitemShut {NoStop}%
\bibitem [{\citenamefont {Foster}\ \emph {et~al.}(2013)\citenamefont {Foster},
  \citenamefont {Dzero}, \citenamefont {Gurarie},\ and\ \citenamefont
  {Yuzbashyan}}]{foster_quantum_2013}%
  \BibitemOpen
  \bibfield  {author} {\bibinfo {author} {\bibfnamefont {M.~S.}\ \bibnamefont
  {Foster}}, \bibinfo {author} {\bibfnamefont {M.}~\bibnamefont {Dzero}},
  \bibinfo {author} {\bibfnamefont {V.}~\bibnamefont {Gurarie}},\ and\ \bibinfo
  {author} {\bibfnamefont {E.~A.}\ \bibnamefont {Yuzbashyan}},\ }\bibfield
  {title} {\bibinfo {title} {Quantum quench in a $p+ip$ superfluid: {Winding}
  numbers and topological states far from equilibrium},\ }\href
  {https://doi.org/10.1103/PhysRevB.88.104511} {\bibfield  {journal} {\bibinfo
  {journal} {Phys. Rev. B}\ }\textbf {\bibinfo {volume} {88}},\ \bibinfo
  {pages} {104511} (\bibinfo {year} {2013})}\BibitemShut {NoStop}%
\bibitem [{\citenamefont {El~Araby}\ and\ \citenamefont
  {Baeriswyl}(2014)}]{el_araby_order_2014}%
  \BibitemOpen
  \bibfield  {author} {\bibinfo {author} {\bibfnamefont {O.}~\bibnamefont
  {El~Araby}}\ and\ \bibinfo {author} {\bibfnamefont {D.}~\bibnamefont
  {Baeriswyl}},\ }\bibfield  {title} {\bibinfo {title} {Order parameter,
  correlation functions, and fidelity susceptibility for the {BCS} model in the
  thermodynamic limit},\ }\href
  {http://link.aps.org/doi/10.1103/PhysRevB.89.134521} {\bibfield  {journal}
  {\bibinfo  {journal} {Phys. Rev. B}\ }\textbf {\bibinfo {volume} {89}}
  (\bibinfo {year} {2014})}\BibitemShut {NoStop}%
\bibitem [{\citenamefont {James}\ \emph {et~al.}(2018)\citenamefont {James},
  \citenamefont {Konik}, \citenamefont {Lecheminant}, \citenamefont
  {Robinson},\ and\ \citenamefont {Tsvelik}}]{james_non-perturbative_2018}%
  \BibitemOpen
  \bibfield  {author} {\bibinfo {author} {\bibfnamefont {A.~J.~A.}\
  \bibnamefont {James}}, \bibinfo {author} {\bibfnamefont {R.~M.}\ \bibnamefont
  {Konik}}, \bibinfo {author} {\bibfnamefont {P.}~\bibnamefont {Lecheminant}},
  \bibinfo {author} {\bibfnamefont {N.~J.}\ \bibnamefont {Robinson}},\ and\
  \bibinfo {author} {\bibfnamefont {A.~M.}\ \bibnamefont {Tsvelik}},\
  }\bibfield  {title} {\bibinfo {title} {Non-perturbative methodologies for
  low-dimensional strongly-correlated systems: {From} non-{Abelian}
  bosonization to truncated spectrum methods},\ }\href
  {https://doi.org/10.1088/1361-6633/aa91ea} {\bibfield  {journal} {\bibinfo
  {journal} {Rep. Prog. Phys.}\ }\textbf {\bibinfo {volume} {81}},\ \bibinfo
  {pages} {046002} (\bibinfo {year} {2018})}\BibitemShut {NoStop}%
\bibitem [{\citenamefont {Robinson}\ \emph {et~al.}(2019)\citenamefont
  {Robinson}, \citenamefont {de~Klerk},\ and\ \citenamefont
  {Caux}}]{robinson_computing_2019}%
  \BibitemOpen
  \bibfield  {author} {\bibinfo {author} {\bibfnamefont {N.~J.}\ \bibnamefont
  {Robinson}}, \bibinfo {author} {\bibfnamefont {A.~J. J.~M.}\ \bibnamefont
  {de~Klerk}},\ and\ \bibinfo {author} {\bibfnamefont {J.-S.}\ \bibnamefont
  {Caux}},\ }\bibfield  {title} {\bibinfo {title} {On computing non-equilibrium
  dynamics following a quench},\ }\href {http://arxiv.org/abs/1911.11101}
  {\bibfield  {journal} {\bibinfo  {journal} {arXiv:1911.11101 [cond-mat]}\ }
  (\bibinfo {year} {2019})}\BibitemShut {NoStop}%
\end{thebibliography}%

\end{document}